\documentclass[pra,twocolumn,floatfix,showpacs,superscriptaddress,groupedaddress]{revtex4}
\usepackage{amsmath,amssymb,amsfonts,bbm,graphicx,times,psfrag}

\usepackage{amsthm}
\usepackage{csquotes}
\usepackage{amsthm}\newcommand{\be} {\begin{equation}}
\newcommand{\ee} {\end{equation}}

\newcommand{\ga}{g_{1}}
\newcommand{\gr}{g_{2}}
\newcommand{\g}{g_{j}}
\usepackage{color}
\renewcommand{\prl}{{Phys. Rev. Lett.} }
\renewcommand{\pra}{{Phys. Rev. A} }

\newcommand{\bra}[1]{\left\langle #1 \right\vert}
\newcommand{\ket}[1]{\left\vert #1 \right\rangle}

\begin{document}
\title{Cavity-induced mirror-mirror entanglement in single-atom Raman laser}
\author{Berihu Teklu}\email{berihut@gmail.com}
\affiliation{Quantum Computing Research Group, Applied Math and Sciences,
Khalifa University, Abu Dhabi, UAE}
\affiliation{New York University Shanghai, 1555 Century Avenue, Pudong, Shanghai 200122, China}
\author{Tim Byrnes}
\affiliation{New York University Shanghai, 1555 Century Avenue, Pudong, Shanghai 200122, China}
\affiliation{State Key Laboratory of Precision Spectroscopy, School of Physical and Material Sciences,
East China Normal University, Shanghai 200062, China}
\affiliation{NYU-ECNU Institute of Physics at NYU Shanghai, 3663 Zhongshan Road North, Shanghai 200062, China}
\affiliation{National Institute of Informatics, 2-1-2 Hitotsubashi, Chiyoda-ku, Tokyo 101-8430, Japan}
\affiliation{Department of Physics, New York University, New York, NY 10003, USA}
\author{Faisal Shah Khan}
\affiliation{Quantum Computing Research Group, Applied Math and Sciences,
Khalifa University, Abu Dhabi, UAE}

\date{\today}

\begin{abstract}
We address an experimental scheme to analyze the optical bistability and the entanglement of two movable mirrors coupled to a two-mode laser inside a doubly resonant cavity. With this aim we investigate the master equations of the atom-cavity subsystem in conjuction with the quantum Langevin equations that describe the interaction of the mirror-cavity. The parametric amplification-type coupling induced by the two-photon coherence on the optical bistabilty of the intracavity mean photon numbers is found and investigated. Under this condition, the optical intensities exhibit bistability for all large values of cavity laser detuning. We also provide numerical evidence for the generation of strong entanglement between the movable mirrors and show that it is robust against environmental thermalization.  

\end{abstract}
\pacs{42.50.Wk, 07.10.Cm, 42.50.Ex, 85.85.+j}
\maketitle

\section{Introduction}
Optomechanics explores the interaction between light and mechanical objects, with a strong experimental focus on micro- and nanoscale systems. It has potential to observe quantum effects like entanglement on a macroscopic object and to apply them to quantum information processing. Furthermore, it may provide a new paradigm for quantum metrology, precision measurement and non-linear dynamical systems. Quantum effects in optomechanical systems (OMS) led to the  demonstration that non-classical states can be generated in an optical cavity \cite{Bose,Mancini}. Moreover, the entanglement between a cavity mode and a mechanical oscillator has been studied both in the steady state \cite{Vitaliprl,Genes} and in the time domain \cite{Mari09,Jie}. In the case of hybrid OMS, the bipartite entanglement between an atomic ensemble, cavity modes and a mirror \cite{Genes08} has shown that a strongly coupled system showing robust tripartite entanglement which can be realized in continuous variable (CV) quantum interfaces \cite{Ian}.

Several schemes have been proposed to generate entanglement between a pair of oscillators interacting with a common bath or in a two-cavity OMS \cite{Paz,Liao}. The long-lived entanglement between two membranes inside a cavity has been studied \cite{Michael} for two mirrors coupled to a cavity \cite{Ling,Ge13a,Ge13,Mancini02}. Several other works have used atomic coherence to induce entanglement \cite{Mancini,Ge13,Ling}. At this point, we should note that entanglement in microcavities is by no means restricted to Raman lasers. As a matter of fact, the formalism can be extended, e.g. to the intersubband case \cite{Auer}, where a strong interplay between photons and the cavity can play an important role for both fundamental and applied physics in the THz to mid Infrared range \cite{Mauro14,Mauro15,Mauro16,Mauro17}.

There have been several quantum features of OMS investigating the generation of macroscopic entangled states for cavity fields due to the atomic coherence in a two-mode laser \cite{Xiong,Kiffner,Qamar,Qamar1,eyob15}, e.g.  the generation of two-mode entangled radiation in a three-level cascade atomic medium \cite{Xiong,eyob15}, a four-level single atom \cite{Kiffner}, and a four-level Raman-driven quantum-beat laser \cite{Qamar}. Entanglement of nanomechanical oscillators and two-mode fields has been achieved via radiation pressure coupling in a cascade configuration due to microscopic atomic coherence. In Ref. \cite{Ling}, two macroscopic mirrors were entangled via microscopic atomic coherence injected into the cavity and Ref. \cite{Ge13} proposed an additional scheme is proposed for entangling two-mode fields whose entanglement can be transferred to two movable mirrors through radiation pressure in a controlled emission laser.  The two photon coherence is generated by strong external classical fields which couples the same levels (dipole-forbidden) in the cascade scheme.

In this paper, we consider a scheme for entanglement generation of two micromechanical mirrors in a four-level atoms in an N configuration through two-mode fields generated by a correlated laser source in a doubly resonant cavity.  All the transitions of interest are dipole allowed. We take the initial state of a four-level atom to be a coherent superposition between of one of the two lower and upper atomic levels, respectively. We show that, in contrast to previous work to the usual the $S$-shaped bistability observed in single-mode optomechanics \cite{Tredicucci,Dorsel,Aspelmeyer}, our scheme shows that the optical intensities of the two cavity modes exhibit bistabilities for all values of detuning, due to the parametric amplification-type coupling induced by the two-photon coherence. We also studied the entanglement created between two movable mirrors in the adiabatic regime and our scheme can control the degree of entanglement with an external field driving the gain medium.  

This paper is organized as follows. The scheme and the Hamiltonian of the system are introduced in Sec. \ref{sec:2}.    In Sec. \ref{sec:3} we analyze the bistability and entanglement between the movable mirrors, by means of a  master equation for the two-mode laser coupled to thermal reservoirs. In Sec \ref{sec:4} we derive the linearized quantum Langevin equations for the field-mirror subsystem. In Sec. \ref{sec:5} we study the coupling induced by the two photon coherence on the bistability of the mean intracavity photon numbers for both cases (RWA and BRWA). We present a method in Sec. \ref{sec:6} to study the entanglement generation between two mirrors using a full numerical analysis of the system. Concluding remarks are given in Sec. \ref{sec:7}.

\begin{figure}[t]
\includegraphics[width=8cm]{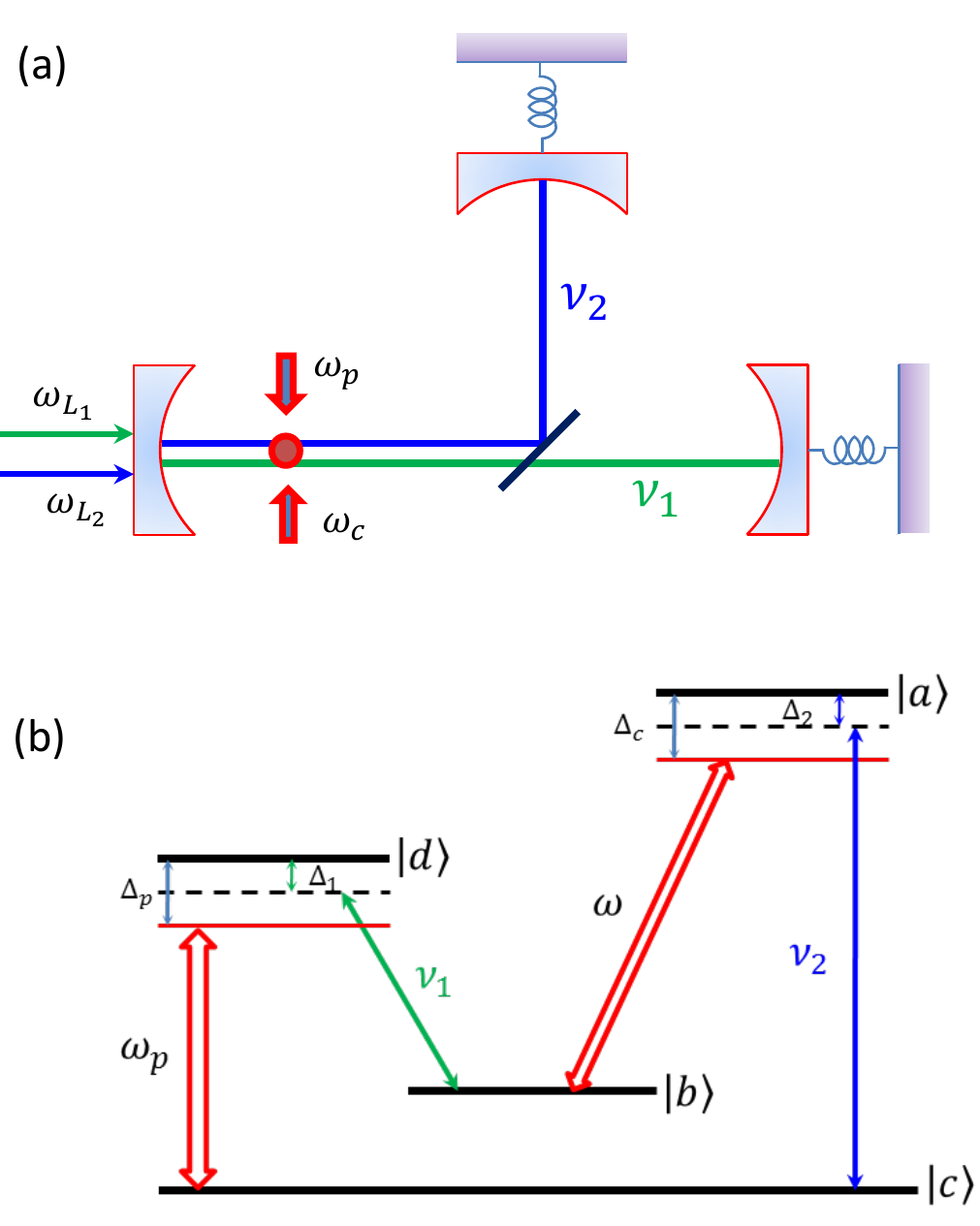}
\caption{(a) Schematics of a two-mode laser coupled to two movable mirrors $M_{1}$ and $M_{2}$. The doubly-resonant cavity is driven by two external lasers of frequency $\omega_{L_{1}}$ and $\omega_{L_{2}}$, and the cavity modes, filtered by a beam splitter (BS), are coupled to their respective movable mirrors by radiation pressure. (b) The gain medium is a single Raman atom. Two external laser drives of frequencies $\omega_{\rm p}$ and $\omega$ are also applied to generate two-photon coherence.}\label{fig1}
\end{figure}
\section{Model of the system}
\label{sec:bigsec2}
\subsection{Hamiltonian}
\label{sec:2}

The OMS which we consider consists of a Fabry-P\'erot cavity of length $L$ with two movable mirrors driven by two mode coherent fields as shown in Fig. \ref{fig1}(a). The laser system is consists of a gain medium of four level atoms in a N configuration as shown in Fig. \ref{fig1}(b). We take the initial state of a four-level atom to be in a coherent superposition of either the state between $\ket a$ and $\ket c$ or $\ket d$ and $\ket b$.  Moreover, a driven laser with amplitude $\Omega_{p}$ and frequency $\omega_{p}$ couple the levels $\ket d$ and $\ket c$ and another laser with amplitude $\Omega$ and frequency $\omega$ couple the levels $\ket a$ and $\ket b$. The atoms are injected into the doubly resonant cavity at a rate $r_a$ and removed after time $\tau$, which is longer than the spontaneous emission time. For the purpose of this paper we take the initial state between $\ket a$ and $\ket c$. The two cavity modes with frequencies $\nu_1$ and $\nu_2$ interact nonresonantly with each atom. We consider the movable mirrors as a quantum harmonic oscillators, so the system can be described by the following Hamiltonian ($\hbar =1$)

\begin{align}
\label{eq:Ham}
H&=\sum_{j=a,b,c,d}\omega_{j}|j\rangle\langle j|+\sum_{j=1}^{2}\nu_{j}a_{j}^{\dag}a_{j}\notag\\
&+g_{1}(a_{1}^{\dag}|b\rangle\langle d|+a_{1}|d\rangle \langle b|)+g_{2}(a_{2}^{\dag}|c\rangle \langle a|+a_{2}|a\rangle \langle c|)\notag\\
&+\Omega_{p}(|d\rangle\langle c|e^{-i\omega_{p}t}+\text{H.c.})+\Omega(|a\rangle\langle b|e^{-i\omega t}+\text{H.c.})\notag\\
&+\sum_{j=1}^{2}[\omega_{\rm m_{j}}b_{j}^{\dag}b_{j}+G_{j}a_{j}^{\dag}a_{j}(b_{j}+b_{j}^{\dag})]\notag\\
&+i\sum_{j=1}^{2}(\varepsilon_{j}a_{j}^{\dag}e^{-i\omega_{\rm L_{j}}t}-\text{H.c.})
\end{align}
where $\omega_{j}$ is the frequency of the $j$th level, $\nu_{j}$ is the frequency of the  $ j $th cavity mode, $g_{j}$ is the atom-field coupling, $\Omega_{p}$ and $\Omega$ are the amplitudes of the drive lasers that couple the $|c\rangle\rightarrow |d\rangle$ and $|b\rangle\rightarrow |a\rangle$ transitions respectively, and $\omega_{p}$, $\omega$ are the frequencies of the drive lasers. $\omega_{\rm m_{j}}$ are the frequencies of the movable mirrors, $b_{j}(b^{\dagger}_{j})$ are the annihilation (creation) operators for the mechanical modes and the relation $G_{j}=(\nu_i/L_j)\sqrt{\hbar/m_{j}\omega_{m_j}}$ is the optomechanical coupling strength, $\left|\varepsilon_{j}\right|=\sqrt{\kappa_j P_j/\hbar\omega_{\rm L_{j}}}$ is the amplitude of the external pump field that drive the doubly resonant cavity, with $\kappa_j$ , $P_j$, and $\omega_{\rm L_{j}}$ being the cavity decay rate related with outgoing modes, the external pump field power and the frequencies of the pump laser, respectively. In (\ref{eq:Ham}), the first two terms denote the free energy of the atom and the cavity modes, the third and the fourth terms represent the atom-cavity mode interaction, the fifth and sixth terms describe the coupling of the levels $|d\rangle \leftrightarrow |c\rangle$ and $|a\rangle \leftrightarrow |b\rangle$  by the drive laser  and the last three terms describe the free energy of mechanical oscillators, the coupling of the mirrors with the cavity modes and the coupling of the external laser derives with the cavity modes, respectively.       

Using the fact that $\sum_{j}|j\rangle\langle j|=1$, the Hamiltonian (\ref{eq:Ham}) can be rewritten (after dropping the unimportant constant $\omega_{b}$) as $H = H_{0}+ H_{I}$ where
\begin{align}
H_{0}&=\omega|a\rangle\langle a|+(\omega-\nu_{2})|c\rangle\langle c|+\nu_{1}|d\rangle\langle d|\notag\\
&+\nu_{1}a_{1}^{\dag}a_{1}+\nu_{2}a_{2}^{\dag}a_{2}
\end{align}
Now applying the transformation $\exp(iH_{0}t)H_{I}\exp(-iH_{0}t)$, we obtain the Hamiltonian in the interaction picture as the sum of the following terms
\begin{align}
\label{eq:cavity-fields}
V_{1}&=\Delta_{c}|a\rangle\langle a|+(\Delta_{c}-\Delta_{2})|c\rangle\langle c|+\Delta_{1}|d\rangle\langle d|\nonumber \\
&+g_{1}(a_{1}^{\dag}|b\rangle\langle d|+a_{1}|d\rangle \langle b|)+g_{2}(a_{2}^{\dag}|c\rangle \langle a|+a_{2}|a\rangle \langle c|)\nonumber \\
&+\Omega_{p}(|d\rangle\langle c|+\text{H.c.})+\Omega(|a\rangle\langle b|+\text{H.c.})\\
V_{2}&=\sum_{j=1}^{2}[\omega_{\rm m_{j}}b_{j}^{\dag}b_{j}+G_{j}a_{j}^{\dag}a_{j}(b_{j}+b_{j}^{\dag})]\nonumber \\
&+i\sum_{j=1}^{2}(\varepsilon_{j}a_{j}^{\dag}e^{i\delta_{j}t}-\text{H.c.})\label{eq:opt1}
\end{align}
where we have assumed for simplicity $\nu_{1}+\nu_{2}=\omega_{p}+\omega$ and we define $\delta_{j}=\nu_{j}-\omega_{\rm L_{j}}$. The master equation for the laser system can be derived using the terms that involve the atomic states $V_{1}$ following the standard laser theory methods \cite{Scu-book97}. In order to obtain the reduced master equation for the cavity modes, it is convenient to trace out the atomic states.

\subsection{Master equation}

\label{sec:3}

In order to obtain the dynamics of the system we require the master equation corresponding to the Hamiltonian (\ref{eq:cavity-fields}). Our model for this system is similar to many earlier treatments of a two-mode three-level laser \cite{Scu-book97,eyob11,eyob15}.  Following these works, we give a derivation of the master equation for our case (proof can be found in the Appendix \ref{app:master}), here we just show the main result.   Assuming that the atoms are injected into the cavity at a rate $r_a$, we can write the density matrix for the atomic and cavity system $\rho_{AR}$ at time $t$ as the sum of the density operator of the cavity modes plus a single atom injected at an earlier time multiplied by the total number of atoms in the cavity for an interval $\Delta t$. Taking the continuum limit, assuming that the atoms are uncorrelated with the electromagnetic modes at the time the atoms are injected into the cavity and when they are removed, and tracing out the atomic degrees of freedom, we obtain the following equation for the density matrix of the cavity modes:
\begin{align}\label{eq:master_eq} 
 \frac{d}{dt}\rho(t)=&-i\ga\left(a_{1}^{\dagger}\rho_{db}+a_{1}\rho_{bd}-\rho_{bd}a_{1}-\rho_{db}a_{1}^{\dagger}\right)\notag\\
 &-i\gr\left(a_{2}\rho_{ca}+a_{2}^{\dagger}\rho_{ac}-\rho_{ac}a_{2}^{\dagger}-\rho_{ca}a_{2}\right)\notag\\
 &+\kappa_1\mathcal{L}[a_1]\rho+\kappa_2\mathcal{L}[a_2]\rho
\end{align}
The last terms in (\ref{eq:master_eq}) are the Lindblad dissipation terms \cite{walls}, where $\kappa_{j}$ are the cavity damping rates added to account for the coupling of the cavity modes with thermal Markovian reservoirs. The conditional density operators $\rho_{ac}={\bra a}\rho_{AR}\ket{c}$ and $\rho_{db}=\bra d\rho_{AR}{\ket b}$ can be obtained from the master equation of the atomic and cavity system, resulting in
\begin{align}
\label{eq:rhoac&bd}
\frac{d}{dt} \rho_{ac}(t)&=-(\gamma_{ac}+i\Delta_2)\rho_{ac}-i\gr(a_2\rho_{cc}-\rho_{aa}a_2)\notag\\
& -i(\Omega\rho_{bc}-\Omega_p\rho_{ad})
 \end{align} 
 \begin{align}
\label{eq:rhoac&bd1}
\frac{d}{dt} \rho_{bd}(t)&=-(\gamma_{bd}-i\Delta_1)\rho_{bd}+i\ga(\rho_{bb}a_{1}^{\dagger}-a_{1}^{\dagger}\rho_{dd})\notag\\
& -i(\Omega\rho_{ad}-\Omega_p\rho_{bc})
 \end{align} 
where $\gamma_{ac}$ and $\gamma_{bd}$ are the dephasing rates for the off-diagonal density matrix elements. We make use of the linear approximation by keeping terms only up to second order in the coupling constants $\g$ in the master equation. This is justified because the coupling constants of the two quantum fields are small as compared to other system parameters \cite{kiffner}. After obtaining the zeroth-order dynamical equations for the conditional density operators other than $\rho_{ac}$ and  $\rho_{bd}$, the good-cavity limit is applied, where the cavity damping rate is much smaller than the dephasing and spontaneous emission rates. The cavity variables then vary more slowly than the atomic ones. The atomic variables reach the steady state earlier than the cavity ones, so we can set the time derivatives of the aforementioned conditioned density operators to zero, being able to solve the system of equations analytically (see Appendix \ref{app:master}).

Here we consider the case in which the atoms are injected into the cavity in a coherent superposition of the levels $|a\rangle$ and $|c\rangle$. We take the initial state as $\ket {\Psi_{A}(0)}={C_{a}(0)}{\ket a}+C_{c}(0){\ket c}$, so the initial density operator for a single atom has the form
\begin{align}\label{eq:superpos}    
\rho_{A}(0)=&\ket \Psi_{A}\bra\Psi_A=\rho^{(0)}_{aa}\ket a\bra a+\rho^{(0)}_{cc}\ket c\bra c\notag\\
+&(\rho^{(0)}_{ac}\ket a\bra c+ \text{H.c}),
\end{align}   
where $\rho^{(0)}_{aa}=|C_a|^2$, $\rho^{(0)}_{cc}=|C_{c}|^2$, and $\rho^{(0)}_{ca}=C_{c}C_{a}^*$ are the initial two level atomic coherence, which can lead to squeezing and entanglement accompanying light amplification \cite{Xiong,Kiffner,eyob07,eyob07a,eyob08,Ge}. It is convenient to introduce the quantity $\eta\in\left[-1,1\right]$ to parametrize the initial density matrix as $\rho_{aa}^{(0)}=\frac{1-\eta}{2}$, and we set the initial coherence to $\rho_{ac}{(0)}=\frac{1}{2}(1-\eta^2)^{1/2}$. The equations of motion, (\ref{eq:rhoac&bd}) and (\ref{eq:rhoac&bd1}) can be solved using the adiabatic approximation, so that the following equations are obtained:
 \begin{align}
 \label{eq:coupled1} 
-i\ga\rho_{bd}=\xi_{6}a_{2}\rho-\xi_5\rho a_{2}+\xi_2\rho a_{1}^{\dagger}-\xi_1a_{1}^{\dagger}\rho\\
-i\gr\rho_{ac}=-\xi_{3}a_{2}\rho+\xi_4\rho a_{2}-\xi_7\rho a_{1}^{\dagger}+\xi_8a_{1}^{\dagger}\rho
 \end{align}
The explicit expressions for the coefficients $\xi_{i}$ can be found in Appendix \ref{app:master}. Finally, the master equation for the cavity modes takes the form 
\begin{align}
\label{eq:master_final} 
 \frac{d}{dt}\rho(t)&=\xi_{1}(a_{1}^{\dagger}\rho a_1-a_1a_{1}^{\dagger}\rho)+\xi_{1}^{*}(a_{1}^{\dagger}\rho a_1-\rho a_1a_{1}^{\dagger})\notag\\
 &+\xi_{2}(a_1\rho a_{1}^{\dagger}-\rho a_{1}^{\dagger}a_1)+\xi_{2}^{*}(a_1\rho a_{1}^{\dagger}-a_{1}^{\dagger}a_1\rho)\notag\\
 &+\xi_{3}(a_2\rho a_{2}^{\dagger}-a_{2}^{\dagger}a_2\rho)+\xi_{3}^{*}(a_2\rho a_{2}^{\dagger}-\rho a_{2}^{\dagger}a_2 )\notag\\
 &+\xi_{4}(a_{2}^{\dagger}\rho a_2-\rho a_2 a_{2}^{\dagger})+\xi_{4}^{*}(a_2^{\dagger}\rho a_{2}-a_{2} a_{2}^{\dagger}\rho)\notag\\
&+\xi_{5}(\rho a_{2} a_{1}-a_{1}\rho a_2)+\xi_{5}^{*}(a_{1}^{\dagger}a_{2}^{\dagger}\rho-a_{2}^{\dagger}\rho a_{1}^{\dagger})\notag\\
 &+\xi_{6}(a_{1} a_{2}\rho-a_{2} \rho a_{1})+\xi_{6}^{*}(\rho a^{\dagger}_{2} a_{1}^{\dagger}-a_{1}^{\dagger}\rho a_{2}^{\dagger})\notag\\
 &+\xi_{7}(\rho a_{1}^{\dagger} a_{2}^{\dagger}-a_{2}^{\dagger}\rho a_{1}^{\dagger})+\xi_{7}^{*}(a_{2} a_{1}\rho-a_{1}\rho a_{2})\notag\\
 &+\xi_{8}(a_{2}^{\dagger} a_{1}^{\dagger}\rho-a_{1}^{\dagger}\rho a_{2}^{\dagger})+\xi_{8}^{*}(\rho a_{1} a_{2}-a_{2}\rho a_{1})\notag\\
 &+\frac{1}{2}\sum_{i=1}^{2}\kappa_{i}[(N_{i}+1)(2a_{i}\rho a_{i}^{\dagger}-a_{i}^{\dagger}a_{i}\rho-\rho a_{i}^{\dagger}a_{i})\notag\\
 &+N_{i}(2a_{i}^{\dagger} \rho a_{i}-a_{i}a_{i}^{\dagger}\rho-\rho a_{i}a_{i}^{\dagger})]  ,
  \end{align}
where we have included the damping of the cavity modes by two independent thermal reservoirs with mean photon number $N_{i}$.

\subsection{Heisenberg-Langevin formulation}

\label{sec:4}

	In order to study the entanglement between the two movable mirrors, we need the quantum Langevin equations for the cavity modes and the mechanical system. Including the creation and annihilation operators for the mechanical system in the Hamiltonian $V_{2}$ and making use of the expression $\left\langle \dot{\mathcal{O}}\right\rangle =Tr\left(\mathcal{O}\dot{\rho}\right)$ we obtain
\begin{align}
\dot a_{1}&=-\bigg(\frac{\kappa_1}{2}-\xi_{11}\bigg)a_1+\xi_{12}a^{\dagger}_{2}\notag\\
&-iG_{1}a_1(b_{1}^{\dagger}+b_1)+\varepsilon_{1}e^{i\delta_{1}t}+F_1 \label{eq:stability}  \\
\dot a_{2}&=-\bigg(\frac{\kappa_2}{2}+\xi_{22}\bigg)a_2-\xi_{21}a^{\dagger}_{1}\notag\\
&-iG_{2}a_2(b_{2}^{\dagger}+b_2)+\varepsilon_{2}e^{i\delta_{2}t}+F_2 \label{eq:stability1a} \\
\dot b_{j}&=-i\omega_{m_j}b_j-\frac{\gamma_{m_j}}{2}b_j-iG_{j}a_{j}^{\dagger}a_j+\sqrt{\gamma_{m_j}}f_{j} \label{eq:stability1}
\end{align} 
where $\xi_{11}=\xi_{1}^{\ast}-\xi_{2}^{\ast}$, $\xi_{12}=\xi_{5}^{\ast}-\xi_{6}^{\ast}$, $\xi_{21}=\xi_{7}-\xi_{8}$ and $\xi_{22}=\xi_{3}-\xi_{4}$. 
The quantum noise operators $F_1$ and $F_2$ appear as a result of the coupling of the external vacuum with the cavity modes and through spontaneous emission. The terms $f_j$ are the noise operators corresponding with a thermal reservoir coupled to the mechanical oscillators.

The quantum noise operators $F_{\mu\nu}$ have zero mean and second-order correlations given by
\begin{equation}
\langle {F_{\mu}(t)} F_{\nu}(t^{\prime})\rangle=2D^{\mu\nu}\delta(t-t^{\prime}).
\end{equation}
Using the generalized Einstein's relations \cite{Cohen,Hald}
\begin{equation}
2\langle D_{\mu\nu}\rangle=-\langle A_{\mu}(t)D_{\nu}(t)\rangle-\langle D_{\mu}(t)A_{\nu}(t)\rangle+\frac{d}{dt}\langle A_{\mu}(t)A_{\nu}(t)\rangle
\end{equation}
where we find that the only nonzero noise correlations between $F_1$, $F_2$, $F_{1}^{\dagger}$ and $F_{2}^{\dagger}$ are 
\begin{align}
\langle {F_{1}^{\dagger}(t)} F_{1}(t^{\prime})\rangle&=2[\text{Re}  (\xi_{1})+\kappa_1N_{1}]\delta(t-t^{\prime}),\\
\langle {F_{1}(t)} F_{1}^{\dagger}(t^{\prime})\rangle&=2[\text{Re} (\xi_{2})+N_{1}(N_{1}+1)]\delta(t-t^{\prime}),\\
\langle {F_{2}^{\dagger}(t)} F_{2}(t^{\prime})\rangle&=2[\text{Re} (\xi_{4})+\kappa_2N_{2}]\delta(t-t^{\prime}),\\
\langle {F_{2}(t)} F_{2}^{\dagger}(t^{\prime})\rangle&=2[\text{Re} (\xi_{3})+\kappa_2(N_{2}+1)]\delta(t-t^{\prime}),\\
\langle {F_{2}(t)} F_{1}(t^{\prime})\rangle&=[\xi_{6}^{*}+\xi_8]\delta(t-t^{\prime}).
\end{align}
Meanwhile, $f_j$ are the noise operators with zero mean contributed by mechanical oscillators and fully characterized by their correlation functions 
\begin{align}
\langle {f_{j}^{\dagger}(t)} f_{j}(t^{\prime})\rangle&=n_{j}\delta(t-t^{\prime}),\\
\langle {f_{j}(t)} f_{j}^{\dagger}(t^{\prime})\rangle&=(n_{j}+1)\delta(t-t^{\prime}),
\end{align}
where $n_j=[\exp(\hbar\omega_{m_{j}}/\kappa_{B}T_j)-1]^{-1}$, is the mean thermal occupation number and $\kappa_{B}$ represents the Boltzmann constant, and $T_j$ is describing the temperature of the reservoir of the mechanical resonator. 

\section{Bistability of intracavity mean photon numbers}\label{sec:5}

\subsection{Mean field expansion}

Typically the single-photon coupling is very weak, but the optomechanical interaction can be greatly enhanced by employing a coherently driven cavity. Bistability has been observed in driven cavity optomechanical systems using a 
Fabry-P\'erot-type optomechanical system in the optical domain \cite{Dorsel,Jiang}. In this section we proceed to study the effect of the coupling induced by the two photon coherence on the bistability of the mean intracavity photon numbers. In order to understand the bistability from the perspective of the intracavity photon number, we consider the steady-state solutions of (\ref{eq:stability})-(\ref{eq:stability1}). This can be performed by transforming the cavity field to its rotating frame, defined by $\tilde{a}_{j}=a_j e^{-i\delta_j t}$, and expanding operators around their mean value:
\begin{align}
\tilde{a}_{j}=\langle \tilde{a}_{j}\rangle+\delta\tilde{a}_{j} \nonumber \\
\tilde{b}_{j}=\langle \tilde{b}_{j}\rangle+\delta\tilde{b}_{j}.
\end{align}
Here, $\langle \tilde{a}_{j}\rangle$ is the average cavity field produced by the laser derive (in the absence of optomechanical coupling), and $\delta\tilde{a}_{j}$ represents the quantum fluctuations around the mean (assumed to be small).  We have also neglected the highly oscillating terms $\exp[-i(\sigma_1+\sigma_2)t]$ in the transformed frame that contains both the fluctuations $\delta\tilde{a}_{j}$ and classical mean values $\langle \tilde{a}_{j}\rangle$. In order to obtain the solutions for $\langle \tilde{a}_{j}\rangle$ in the steady state, one must either simplify the equations by making a rotating wave approximation which neglects the fast oscillating terms, or solve a set of self-consistent equations.  We show both of these approaches in the following subsections.  

\subsection{Rotating wave approximation}

In the Rotating wave approximation (RWA), we neglect fast oscillating terms in the transformed quantum Langevin approach to determine the evolution for $\langle \tilde{b}_{j}\rangle$ and $\langle \tilde{a}_{j}\rangle$. This gives the steady-state solutions according to
\begin{align} 
\langle b^{\dagger}_j+b_j\rangle&=-\frac{2\omega_{m_{j}}G_{j}I_j}{\gamma^{2}_{m_{j}}/4+\omega^2_{m_{j}}},\\
\langle \tilde{a}_{j}\rangle&=\frac{\varepsilon_{j}}{i\delta_j+\kappa_{j}/2+(-1)^j\eta_j},
\label{eq:mean}
\end{align}
where 
\begin{align}
I_j=|\langle \tilde{a}_{j}\rangle|^2
\end{align}
is the steady-state intracavity mean photon number,
\begin{align}
\delta_{j}=\nu_{j}-\omega_{L_{j}}+G_{j}\langle b^{\dagger}_{j}+b_{j}\rangle
\end{align}
is the cavity mode detuning, and we have chosen the frequency shift due to radiation pressure 
\begin{align}
\delta\nu_j\equiv G_{j}\langle b^{\dagger}_j+b_j\rangle
\end{align}
for convenience.  We have also defined
\begin{align}
\eta_{1}& =\xi^{*}_{1}-\xi^{*}_{2} \nonumber \\
\eta_2 & = \xi_{3}-\xi_{4} .
\end{align}
We can then write the equations for the intracavity mean photon numbers to have the implicit form 
\begin{equation}
\label{eq:bistability}
I_j\bigg|i(\delta_{0j}-\beta_{j} I_{j})^2+\frac{\kappa_j}{2}+(-1)^j\eta_{j}\bigg|^2=|\varepsilon_{j}|^2,
\end{equation}
where we have used 
\begin{align}
\delta_{0j}=\nu_{j}-\omega_{L_{j}}
\label{delta0jdef}
\end{align}
and 
\begin{align}
\beta_{j}=(2\omega_{m_{j}}G^2_{j})/(\gamma^2_{m_{j}}/4+\omega^2_{m_j}).
\end{align}
Eq. (\ref{eq:bistability}) is of the form of the standard equations for ${\rm S}$-shaped bistabilities for intracavity intensities in an optomechanical system with effective cavity damping rates $\kappa_j++2(-1)^{j}\eta_j$. We would note that typically in RWA, there is no coupling between the intensities of the cavity modes that is due to the two-photon coherence induced in the system. 

To demonstrate the bistable behavior of the mean intracavity photon numbers in doubly resonant cavity, we use a set of particular parameters from recent available experimental setups \cite{Gr,Ar}. We consider mass of the mirrors $m=145{\rm ng}$, cavity with lengths $L_1= 112{\rm\mu m}$, $L_2=88.6{\rm\mu m}$, pump laser wavelengths $\lambda_1 =810{\rm nm}$, $\lambda_2 =1024{\rm nm}$, rate of injection of atoms $r_a =1.6{\rm MHz}$, mechanical oscillator damping rates $\gamma_{m_{1}} =\gamma_{m_{2}} =2\pi\times 60{\rm MHz}$, mechanical frequencies $\omega_{m_{1}}  =\omega_{m_{2}} =2\pi\times3{\rm MHz}$, and without loss of generality, we assume that the dephasing and spontaneous emission rates for the atoms $\gamma_{ac} =\gamma_{bd} =\gamma_{cd} =\gamma_{ab}=\gamma_{bc} =\gamma_{ad} =\gamma_{a} =\gamma_{b}  =\gamma_{c} =\gamma_d =\gamma  =3.4 {\rm MHz}$. For the purpose of this paper, we assume a Gaussian distribution for the atom density and set both one- and two-photon detunings to $0$, $\Delta_{p} =0$ and $\Delta_{c} =0$, respectively. 

\begin{figure}[tb]
\vspace{0.5cm}
\begin{center}
\includegraphics[width=0.4955\columnwidth,height=2in]{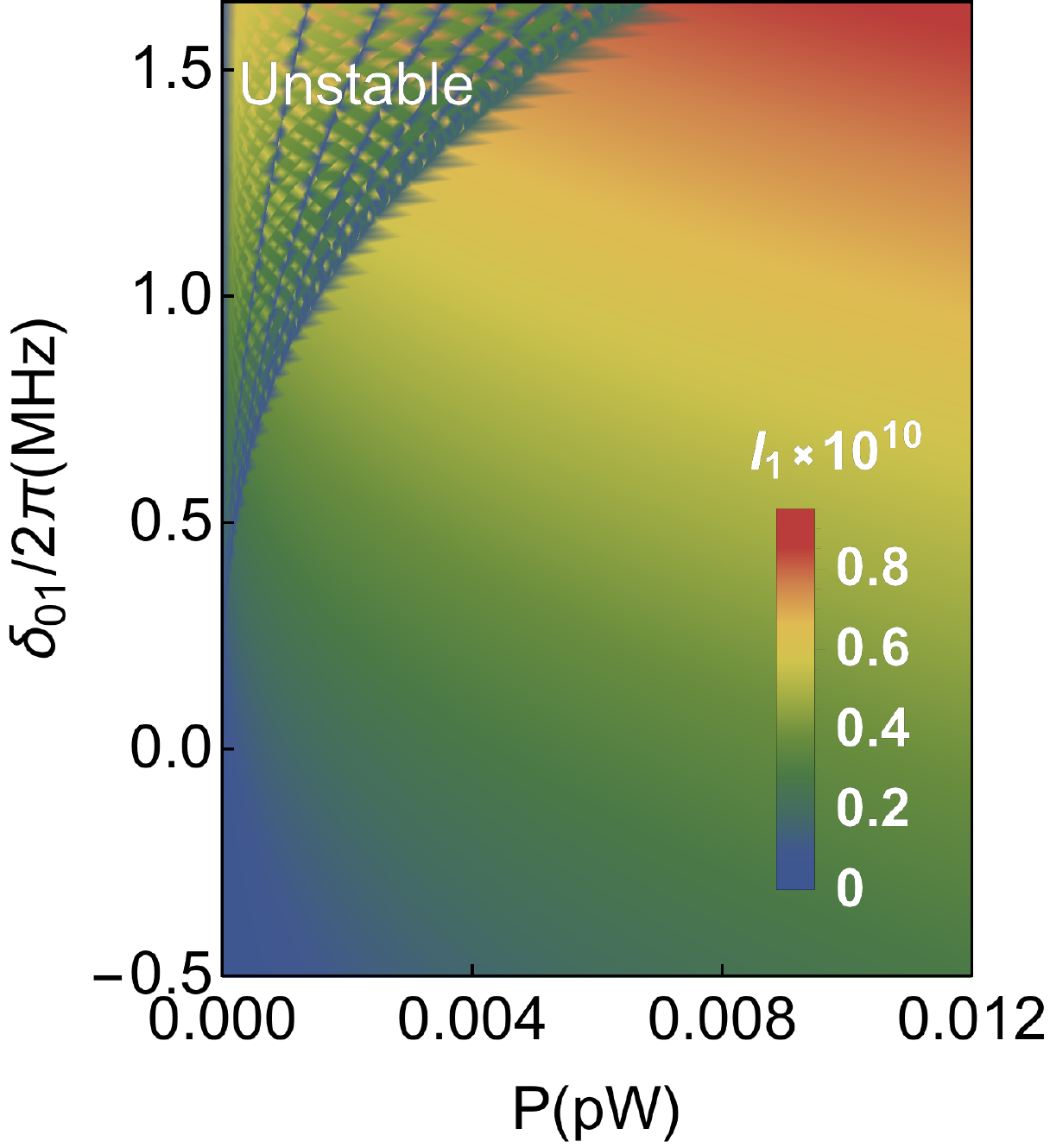}
\includegraphics[width=0.4955\columnwidth,height=2in]{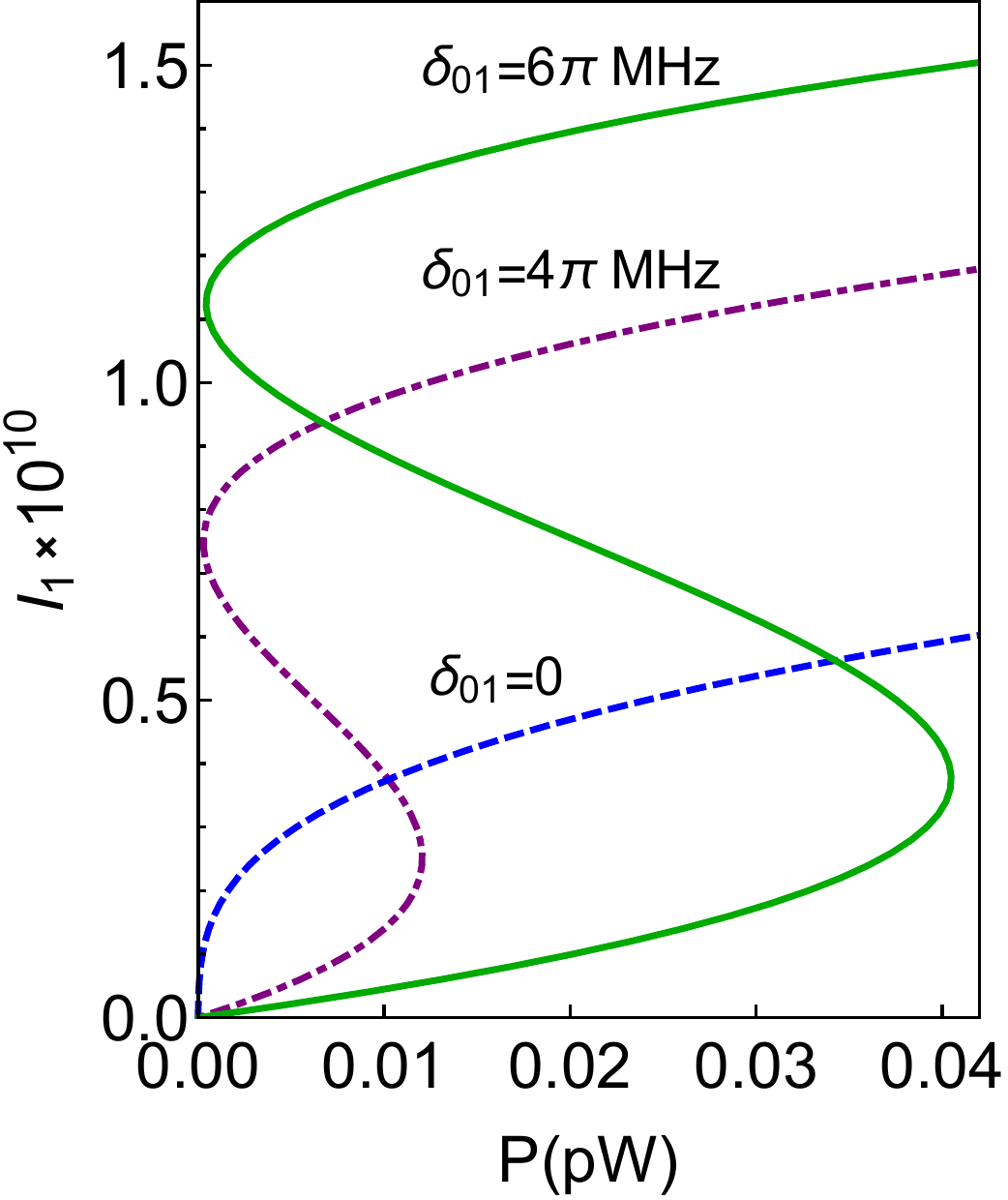}
\end{center}
\vspace{-0.5cm}
 \caption{\label{f:bista}
Tunable optical bistability of intracavity field. The left panel shows the phase diagram for the intra-cavity mean photon number ${\rm I_{1}}$ for different values of cavity laser detuning ${\rm \delta_{01}}$ and external pump field strength (cavity drive laser) $P$ in the rotating wave approximation. The right panel red, green, and blue curves are the cross section of the phase diagram for cavity laser detuning $\delta_{01} =6\pi{\rm MHz}, 4\pi{\rm MHz}, 0$, respectively. Here $g_{1} =g_{2} =2\pi\times3{\rm MHz}$, $\Omega/\gamma =10$, $\kappa_{1} =\kappa_{2} =2\pi \times 215{\rm kHz}$, and assuming that all atoms are initially in their excited state $|\psi_{A}(0)\rangle=\ket{c}$ corresponding to the parameter $\eta=1$.}
\end{figure}

We first illustrate the bistability of the steady-state intracavity mean photon number for the first cavity mode. The first example we present in Fig. \ref{f:bista} is the steady-state intracavity photon number under the red-detuned (${\rm \delta_{01}}>0$) frequency range.  We point out that we have introduced the effective detuning for our system (\ref{delta0jdef}), where the red-detuned regime occurs for all positives values of the effective detuning, which is the opposite to prior conventions \cite{Clerk}. The left panel of Fig. (\ref{f:bista}) shows that the optical bistability regime persists for a broader range of the external pump fields. The right panel shows that an {\rm S}-shaped behavior of the bistable intra-cavity mean photon number for ${\rm I_{1}}$. The strength of the bistability is changed by increasing the intensity of the external field and the detuning. For the second cavity mode,  we have found almost exactly the same results for the bistability behavior of the steady-state intracavity mean photon number. 

\subsection{Beyond the rotating wave approximation}

Let us analyze the bistability behavior of the intracavity mean photon number in the NRWA. In this case we are able to see the effect of the two-photon coherence. To study the bistabilty in the regime, we consider the rotating frame defined by the bare cavity frequencies $\nu_{j}$. This is equivalent to the assumption that the cavity mode detunings $\delta\nu_{j}=0$ in the Hamiltonian (\ref{eq:opt1}).  It stays in the counter-rotating terms in the Langevin equations for $\tilde{a}_{j}$. This approach can be traced back the condition 
\begin{align}
{\rm \delta_{02}}=-{\rm \delta_{01}}\equiv-{\rm \delta_{0}}.
\end{align}
The expectation values of the cavity mode operators with this choice of detuning are
\begin{align} \label{eq:exp1}
\langle \tilde{a}_{1}\rangle&=\frac{\varepsilon_{1}\alpha^{*}_2+\varepsilon_{2}(\xi^{*}_{5}-\xi^{*}_{6})}{\alpha_2\alpha^{*}_2+(\xi^{*}_{5}-\xi^{*}_{6})(\xi^{*}_{7}-\xi^{*}_{8})} \\
\langle \tilde{a}_{2}\rangle&=\frac{\varepsilon_{2}\alpha^{*}_1-\varepsilon_{1}(\xi_{8}-\xi_{7})}{\alpha^{*}_{1}\alpha_{2}+(\xi_{5}-\xi_{6})(\xi_{7}-\xi_{8})},\label{eq:exp2}
\end{align}
where 
\begin{align}
\alpha_1& =i(\delta_{0}-\beta_1I_1)+\kappa_{1}/2-\eta_{1} \nonumber \\
\alpha_2& =-i(\delta_{0}+\beta_{2}I_{2})+\kappa_{2}/2+\eta_{2}.
\end{align}
As can be seen in (\ref{eq:exp1}) and (\ref{eq:exp2}), the coupling between $\langle \tilde{a}_{1}\rangle$ and $\langle \tilde{a}_{2}\rangle$ is due to the coefficients $\xi_7$ and $\xi_8$, which are proportional to the coherence induced either by the coupling of atomic levels by an external laser, or by injecting the atoms in a coherent superposition of upper and lower levels. Here, we consider a more general expression by introducing a new parameter that relates the cavity drive amplitudes  $ (P_{2}\sim\mu^{2}P_{1})$
\begin{equation}
|\varepsilon_{2}|=\mu|\varepsilon_{1}|\equiv\mu|\varepsilon|.
\end{equation}
We thus obtain an equivalent relation for the intracavity mean photon number
\begin{align} \label{eq:exp1a}
\frac{|\alpha_{1}(I_1)\alpha^{*}_{2}(I_{2})+(\xi^{*}_{5}-\xi^{*}_{6})(\xi^{*}_{7}-\xi^{*}_{8})|^2}{|\alpha^{*}_{2}(I_{2})+\mu(\xi^{*}_{5}-\xi^{*}_{6})|^2}I_1=|\varepsilon|^2,\\
\frac{|\alpha^{*}_{1}(I_1)\alpha_{2}(I_{2})+(\xi_{5}-\xi_{6})(\xi_{7}-\xi_{8})|^2}{|\mu\alpha^{*}_{1}(I_{1})-(\xi_{7}-\xi_{8})|^2}I_2=|\varepsilon|^2\label{eq:exp2a}. 
\end{align}

The above transformation provides an elegant approach to understanding the effect of the coupling on the bistability behavior of the cavity modes by examining the limits of the parameter $\mu^2$. In the limit where $\mu^2\ll1$ ($P_2\ll P_1$), the denominator in (\ref{eq:exp2a}) can be approximated as 
\begin{align}
 & |\mu\alpha^{*}_{1}-(\xi_{7}-\xi_{8})|^2 \nonumber \\
& \approx |\mu(-i\delta_0+\kappa_{1}/2-(\xi_{1}-\xi_{2}))-(\xi_{7}-\xi_{8})|^2
\end{align}
for $\mu^2\beta_{1}I_{1}/|(\xi_{7}-\xi_{8})|^2\ll 1$. In this case, the ratio of (\ref{eq:exp1a}) and (\ref{eq:exp2a}) yields a cubic equation
\begin{align}
I_1/I_{2}= \frac{|\alpha^{*}_{2}(I_{2})+\mu(\xi^{*}_{5}-\xi^{*}_{6})|^2}{|\mu(-i\delta_0+\kappa_{1}/2-(\xi^{*}_{1}-\xi^{*}_{2}))-(\xi_{7}-\xi_{8})|^2}.
\end{align}
Note that this implies that $I_{2}$ exhibits bistability when the intensity of the first cavity mode is varied.

\begin{figure}[tb]
\vspace{0.5cm}
\begin{center}
\includegraphics[width=0.45\columnwidth,height=2in]{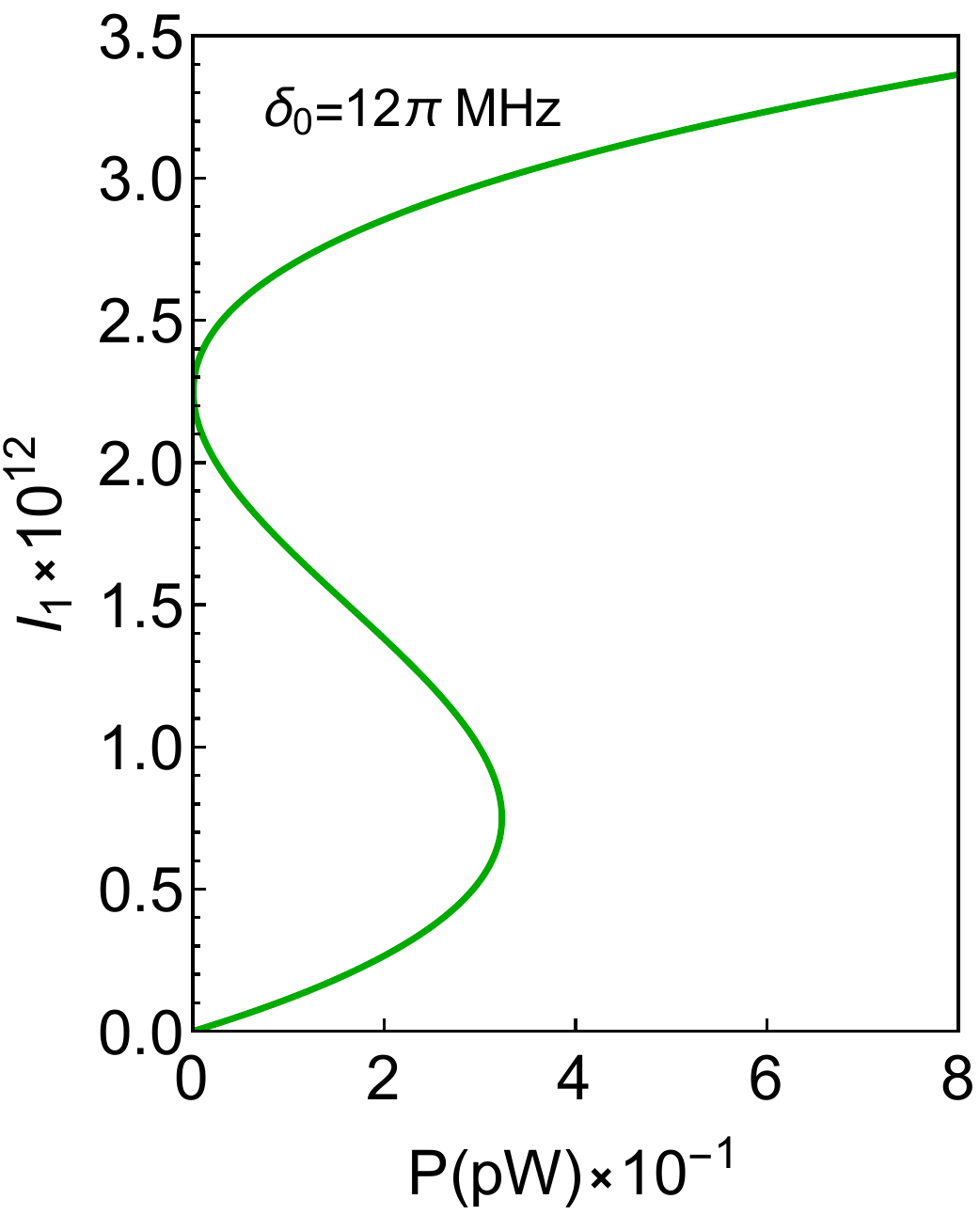}
\includegraphics[width=0.45\columnwidth,height=2in]{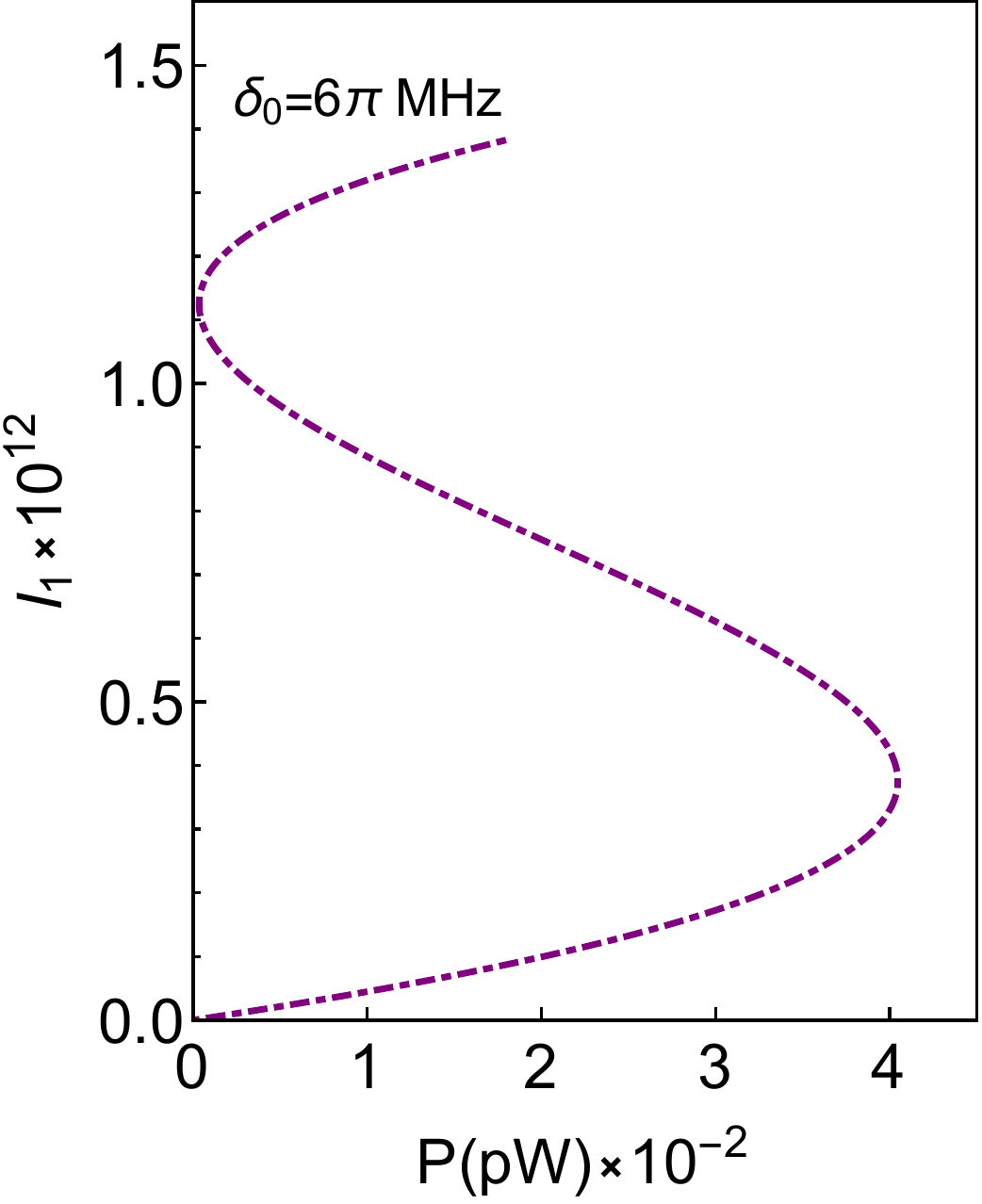}\\
\vspace{0.5cm}
\includegraphics[width=0.45\columnwidth,height=2in]{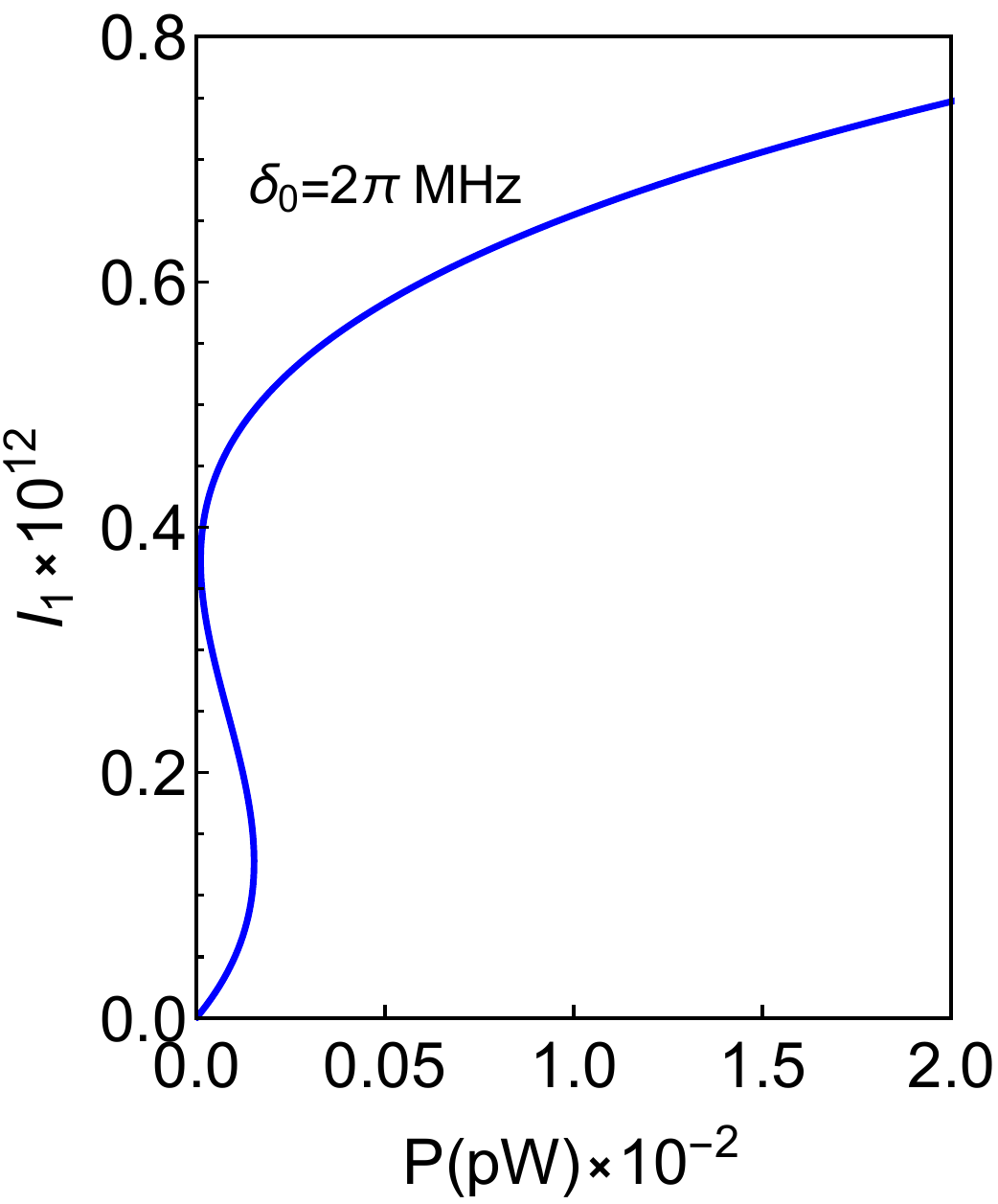}
\includegraphics[width=0.45\columnwidth,height=2in]{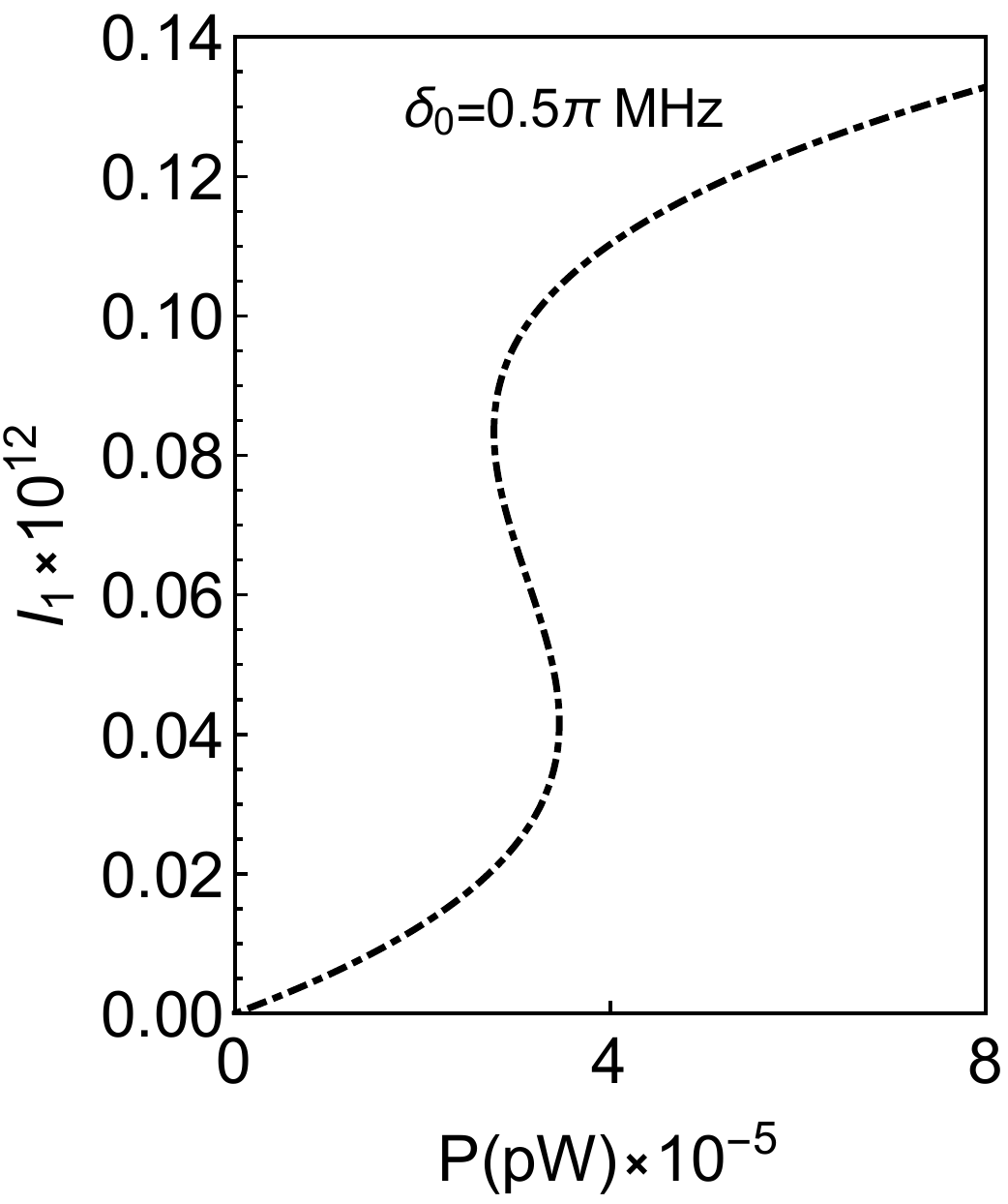}
\end{center}
\vspace{-0.5cm}
 \caption{\label{f:bista1}
Cross section of the phase diagram at $\delta_{0}/2\pi=6{\rm MHz}$, $\delta_{0}/2\pi=3{\rm MHz}$, $\delta_{0}/2\pi=1{\rm MHz}$, and $\delta_{0}/2\pi=0.25{\rm MHz}$. Notice that the bistability appears for positive values of detuning, which good agreement is achieved in the ``red detuned'' regime which allows in single-mode optomechanics \cite{Tredicucci,Dorsel,Aspelmeyer}. Here we have used $\mu=0.1$ ($P_2=0.08P_1$), and atoms are initially injected into the cavity in state $|\psi_{A}(0)\rangle=\ket{c}$, that is, for the value of the parameter $\eta=1$. See text and Fig.  \ref{f:bista} for the other parameters.}
\end{figure}

An exact numerical analysis on Eqs. (\ref{eq:exp1a}) and (\ref{eq:exp2a}) is shown in Fig. (\ref{f:bista1}), which indicates that the behavior of the cavity mode mean photon number is very sensitive to the sign of detuning. As can be seen in the RWA case, the bistabilty occurs in the ``red detuned'' regime ($\delta_{0}>0$)-good agreement is achieved in the regimes of validity of each model. We also observe that the bistable region widens with increasing detuning and derive laser power.
\section{Dynamics of continuous variable entanglement}\label{sec:6}

In this section, we investigate the degree of entanglement of the movable mirrors of the doubly resonant cavity in the adiabatic regime. The detection of entanglement in similar contexts has been 
investigated by many groups recently \cite {Xiong,eyob07,eyob07a,eyob08,Rist,Palo,Wang}. Although there is no entanglement between the cavity fields and the movable mirrors, here we will show that the entanglement between the two-mode fields can be transferred to entanglement between the movable mirrors of the doubly resonant cavity. Indeed, optimal entanglement transfer from the two-mode cavity field to the mechanical modes is achieved by eliminating adiabatically the dynamics of the field modes, specifically in circumstances where $\kappa_j\gg\gamma_{m_{j}}$.   

We introduce the slowly varying fluctuation operators $\delta a_{j}\equiv\delta \tilde{a}_{j}e^{i\delta_{j}t}$ and $\tilde{b}_{j}\equiv b_{j}e^{i\omega_{m_{j}}t}$ and using (\ref{eq:stability})-(\ref{eq:stability1}), the corresponding linear quantum Langevin equations are written 
\begin{align}
\delta \dot {a}_1&=-\frac{\kappa^{\prime}_{1}}{2}\delta a_1+\xi_{12}\delta a^{\dagger}_{2}-i G_{1}\langle \tilde{a}_{1}\rangle(\delta\tilde{b}^{\dagger}_{1}e^{i(\delta_1+\omega_1)t}\notag\\
&+\delta \tilde{b}_{1}e^{i(\delta_1-\omega_1)t})+F_{1}\\ 
\delta \dot {a}_2&=-\frac{\kappa^{\prime}_{2}}{2}\delta a_2-\xi_{21}\delta a^{\dagger}_{1}-i G_{2}\langle\tilde{a}_{2}\rangle(\delta \tilde{b}^{\dagger}_{2}e^{i(\delta_2+\omega_2)t}\notag\\ 
&+\delta \tilde{b}_{2}e^{i(\delta_2-\omega_2)t})+F_{2}\\ 
\delta\dot{\tilde{b}}_{j}&=-\frac{\gamma_{m_{j}}}{2}\delta\tilde{b}_{j}-i G_{j}\left\langle \tilde{a}_{j}\right\rangle \delta a_{j}^{\dagger}e^{i\left(\omega_{m_{j}}+\delta_{j}\right)t}\notag\\ 
&-i G_{j}\left\langle \tilde{a}_{j}^{\dagger}\right\rangle \delta a_{j}e^{i\left(\omega_{m_{j}}-\delta_{j}\right)t}+\sqrt{\gamma_{m_{j}}}f_{j}
\end{align}
where $\kappa'_{1}=\kappa_{1}-2\xi_{11}$, $\kappa'_{2}=\kappa_{2}+2\xi_{22}$. Here we have the choice of using using the RWA when evaluating $ \langle \tilde{a}_{j} \rangle $. In the RWA,
the model should not enter the regime where the measurement is capable of resolving the zero-point motion of the oscillator in a time short compared with the mechanical oscillation period. This regime -- which requires very strong optomechanical coupling -- exhibits interesting behavior, including dynamical mechanical squeezing \cite{Doherty,Warwick}. From the perspective of quantum state transfer, it has been shown in \cite{Pinard,Aspelmeyer} that the optomechanical interaction and consequently the field-mirror entanglement are enhanced when the detuning of each cavity-driving field is $\delta_{j}=-\omega_{m_{j}}$.        
To avoid these issues, we explicitly compute the $ \langle \tilde{a}_{j} \rangle $ without using the RWA by using a self-consistent iterative approach. Setting $\delta_{j}=-\omega_{m_{j}}$ and using the adiabatic approximation for the $\delta a_j$ equations we get the expressions for the mirror variables. Moreover, we can choose the phase of the driving laser in such a way that $\left\langle \tilde{a}_{j}\right\rangle =-i\left|\left\langle \tilde{a}_{j}\right\rangle \right|$.
Hence, we have the final expressions
\begin{align}
\delta\dot{\tilde{b}}_{1}&=-\frac{\gamma_{m_{1}}}{2}\delta\tilde{b}_{1}+\alpha_{1}\left(e^{2i\delta_{2}t}-e^{-2i\delta_{1}t}\right)\delta\tilde{b}_{2}\notag\\
&+\alpha_{1}\left(1-e^{-2i\left(\delta_{1}+\delta_{2}\right)t}\right)\delta\tilde{b}_{2}^{\dagger}+\widetilde{F}_{1},\\
\delta\dot{\tilde{b}}_{2}&=-\frac{\gamma_{m_{2}}}{2}\delta\tilde{b}_{2}+\alpha_{2}\left(e^{-2i\delta_{2}t}-e^{2i\delta_{1}t}\right)\delta\tilde{b}_{1}\notag\\
&+\alpha_{2}\left(e^{-2i\left(\delta_{1}+\delta_{2}\right)t}-1\right)\delta\tilde{b}_{1}^{\dagger}+\widetilde{F}_{2}
\end{align}
where
\begin{align}
\alpha_{1}&\equiv\frac{4\xi_{12}}{\kappa}G_{1}G_{2}\left|\left\langle \tilde{a}_{1}\right\rangle \right|\left|\left\langle \tilde{a}_{2}\right\rangle \right| \notag\\
\widetilde{F}_{1}&\equiv\frac{2\kappa'_{2}}{\kappa}G_{1}\left|\left\langle \tilde{a}_{1}\right\rangle \right|\left(e^{-2i\delta_{1}t}F_{1}-F_{1}^{\dagger}\right)\notag\\
&+\frac{4\xi_{12}}{\kappa}G_{1}\left|\left\langle \tilde{a}_{1}\right\rangle \right|\left(e^{-2i\delta_{1}t}F_{2}^{\dagger}-F_{2}\right)+\sqrt{\gamma_{m_{1}}}f_{1}\notag\\
\alpha_{2}&\equiv\frac{4\xi_{21}}{\kappa}G_{1}G_{2}\left|\left\langle \tilde{a}_{1}\right\rangle \right|\left|\left\langle \tilde{a}_{2}\right\rangle \right|\notag\\
\widetilde{F}_{2}&\equiv\frac{2\kappa'_{1}}{\kappa}G_{2}\left|\left\langle \tilde{a}_{2}\right\rangle \right|\left(e^{-2i\delta_{2}t}F_{2}-F_{2}^{\dagger}\right)\notag\\
&+\frac{4\xi_{21}}{\kappa}G_{2}\left|\left\langle \tilde{a}_{2}\right\rangle \right|\left(F_{1}-e^{-2i\delta_{2}t}F_{1}^{\dagger}\right)+\sqrt{\gamma_{m_{2}}}f_{2}
\end{align}
with $\kappa'_{1}=\kappa_{1}-2\xi_{11}$, $\kappa'_{2}=\kappa_{2}+2\xi_{22}$ and $\kappa=\kappa'_{1}\kappa'_{2}+4\xi_{12}\xi_{21}$.

In order to study the entanglement between the mirrors it is convenient to define the position and momentum operators as $\delta q_{j}=\frac{\delta\tilde{b}_{j}+\delta\tilde{b}_{j}^{\dagger}}{\sqrt{2}}$
and $\delta p_{j}=i\frac{\delta\tilde{b}_{j}^{\dagger}-\delta\tilde{b}_{j}}{\sqrt{2}}$. Once we get the expressions for these fluctuation operators we can write a matrix equation of the form 
\begin{equation}
\label{eq:time_dep}
\dot{\bf u}\left(t\right)={\bf M}\left(t\right){\bf u}\left(t\right)+{\bf n}\left(t\right),
\end{equation}
where we define  
\begin{align}
{\bf u}=\left(\begin{array}{cccc}
\delta q_{1}, & \delta p_{1},& \delta q_{2},& \delta p_{2}\end{array}\right)^{T} .
\end{align}
Here ${\bf M}$ is a matrix containing the coupling between the fluctuations and the vector ${\bf n}$ contains the noise operators of both the cavity and the mirrors. This inhomogeneous  differential equation can be solved numerically. The evolution of the quadrature fluctuations is described by the general solution of (\ref{eq:time_dep}) is formally expressed as \cite{Mari,Mari09,Jie}
\begin{equation}
{\bf u}\left(t\right)=\mathbb{\bf G}\left(t\right){\bf u}\left(0\right)+\mathbb{\bf G}\left(t\right)\int_{0}^{t}\mathbb{\bf G}^{-1}\left(\tau\right){\bf n}\left(\tau\right)d\tau,
\end{equation}
where 
\begin{align}
\mathbb{\bf G}\left(t\right)=e^{\int M\left(s\right)ds}
\end{align}
and the initial condition satisfies $\mathbb{\bf G}\left(0\right)=\mathbb{I}$, where $\mathbb{I}$ is the identity matrix. To bring quantum effects to the macroscopic level, one important way is the creation of entanglement between the optical mode and the mechanical mode. If the initial state of the system is Gaussian, the statistics remain Gaussian under continuous linear measurement for all time. 

The entanglement can therefore be quantified via the logarithmic negativity. The logarithmic negativity is a convenient and commonly used parameter to quantify the strength of a given entanglement resource and has the attractive properties of both being additive for multiple independent entangled states and quantifying the maximum distillable entanglement \cite{Vidal}. Here, we will quantify
the entanglement by means of the logarithmic negativity. In particular, such measurement can be obtained from the correlation matrix $\mathbb{\bf V}$ with elements given by 
\begin{align}
V_{i,j}\equiv\frac{1}{2}\left\langle u_{i}u_{j}+u_{j}u_{i}\right\rangle+\left\langle u_{i}\right\rangle \left\langle u_{j}\right\rangle 
\end{align}
fully characterizes the mechanical and optical variances (It also includes information  on  the  quantum  correlation  between  the  two  mechanical and the optical cavity modes), giving rise to a block structure:
\begin{equation}
\mathbb{\bf V}=\left(\begin{array}{cc}
\mathbb{\bf A} & \mathbb{\bf C}\\
\mathbb{\bf C}^{T} & \mathbb{\bf B}
\end{array}\right)
\end{equation}
The corresponding logarithmic negativity $E_{\mathcal{N}}$ is given by \cite{Adesso,Ferraro}
\begin{equation}
E_{\mathcal{N}}=\max \left( 0, -\ln  2\eta^{-}\right),
\label{entdef}
\end{equation}
where 
\begin{align}
\eta^{-} =\frac{1}{\sqrt{2}}\sqrt{\Sigma-\sqrt{\Sigma^{2}-\det V}} 
\end{align}
is the symplectic eigenvalue with regards to quantum correlations and
\begin{align}
\Sigma & =\text{det}  \mathbb{\bf A}+\det   \mathbb{\bf B}-2 \det \mathbb{\bf C} .
\end{align}
The interesting quantities in the present model are the quadrature fluctuations of the cavity and the mirror. Since the fluctuations are time dependent, so will be the elements of the correlation matrix. In order to compute its elements, we define a covariance matrix $\mathbb{\bf R}\left( t \right)$ by the elements %
\begin{align}
\mathbb{\bf R}_{\ell,\ell^{\prime}}\left(t\right)=\left\langle u_{\ell}\left(t\right)u_{\ell^{\prime}}\left(t\right)\right\rangle
\end{align}
for $\ell,\ell^{\prime}=1, 2,3, 4$.

In order to quantify the two-mode entanglement, we need to determine the covariance matrix $\mathbb{\bf R}\left(t\right)$.     Taking into account the (\ref{eq:time_dep}) and assuming that the correlation between its elements and the noise operators at the initial state is zero, the general expression for the covariance matrix $\mathbb{\bf R}\left(t\right)$ at an arbitrary time has takes the form
\begin{equation}
\mathbb{\bf R}\left(t\right)=\mathbb{\bf G}\left(t\right)
\mathbb{\bf R}\left(0\right)\mathbb{\bf G}^{T}\left(t\right)+\mathbb{\bf G}\left(t\right)\mathbb{\bf Z}\left(t\right)\mathbb{\bf G}^{T}\left(t\right),
\end{equation}
where 
\begin{align}
\mathbb{\bf Z}\left(t\right)=\int_{0}^{t}\int_{0}^{t}\mathbb{\bf G}^{-1}\left(\tau\right)\mathbb{\bf C}\left(\tau,\,\tau'\right)\left[\mathbb{\bf G}^{-1}\left(\tau^{\prime}\right)\right]^{T}d\tau\,d\tau'.
\end{align}
The elements of the matrix $\mathbb{\bf C}\left(\tau,\,\tau'\right)$ are the correlation between the elements
of the vector $\mathbb{\bf n}$, that is, 
\begin{align}
\mathbb{\bf C}_{l,m}\left(\tau,\,\tau'\right)=\left\langle \mathbb{\bf n}_{l}\left(\tau\right)\mathbb{\bf n}_{m}\left(\tau'\right)\right\rangle .
\end{align}
Those elements can be easily calculated by using the generalized Einstein relation for the noise operators. Moreover, since the expectation value for the noise operators is zero, the equation for the mean value of the
fluctuations is 
\begin{align}
\left\langle \mathbb{\bf u}\left(t\right)\right\rangle =\mathbb{\bf G}\left(t\right)\left\langle \mathbb{\bf u}\left(0\right)\right\rangle.
\end{align}

The above are the formal equations for the evolution of quadrature operators of the mirrors. We now assume the density matrix of the initial conditions of the mirrors is separable and the mechanical bath is, as usual, in a thermal state at temperature $T$  with occupancy $\mathbb{\bf n}_{th}$ and the cavity mode is in vacuum state. Therefore, the initial density matrix for the $i$th mechanical oscillator is given by
\begin{equation}
\rho_{m_{i}}=\sum_{j=0}^{\infty}\frac{n_{i}^{j}}{\left(1+n_{i}\right)^{j+1}}\left|j\right\rangle \left\langle j\right|.
\end{equation}
Under this assumption the $\mathbb{\bf R}$ matrix at the initial state is given by
\begin{equation}
\mathbb{\bf R}\left(0\right)=\left(\begin{array}{cccc}
n_{1}+\frac{1}{2} & \frac{i}{2} & 0 & 0\\
-\frac{i}{2} & n_{1}+\frac{1}{2} & 0 & 0\\
0 & 0 & n_{2}+\frac{1}{2} & \frac{i}{2}\\
0 & 0 & -\frac{i}{2} & n_{2}+\frac{1}{2}
\end{array}\right).
\end{equation}
From (\ref{entdef}), the entanglement of the movable mirrors can be easily computed numerically. 

\begin{figure}[tb]
\vspace{0.5cm}
\begin{center}
\includegraphics[width=0.995\columnwidth]{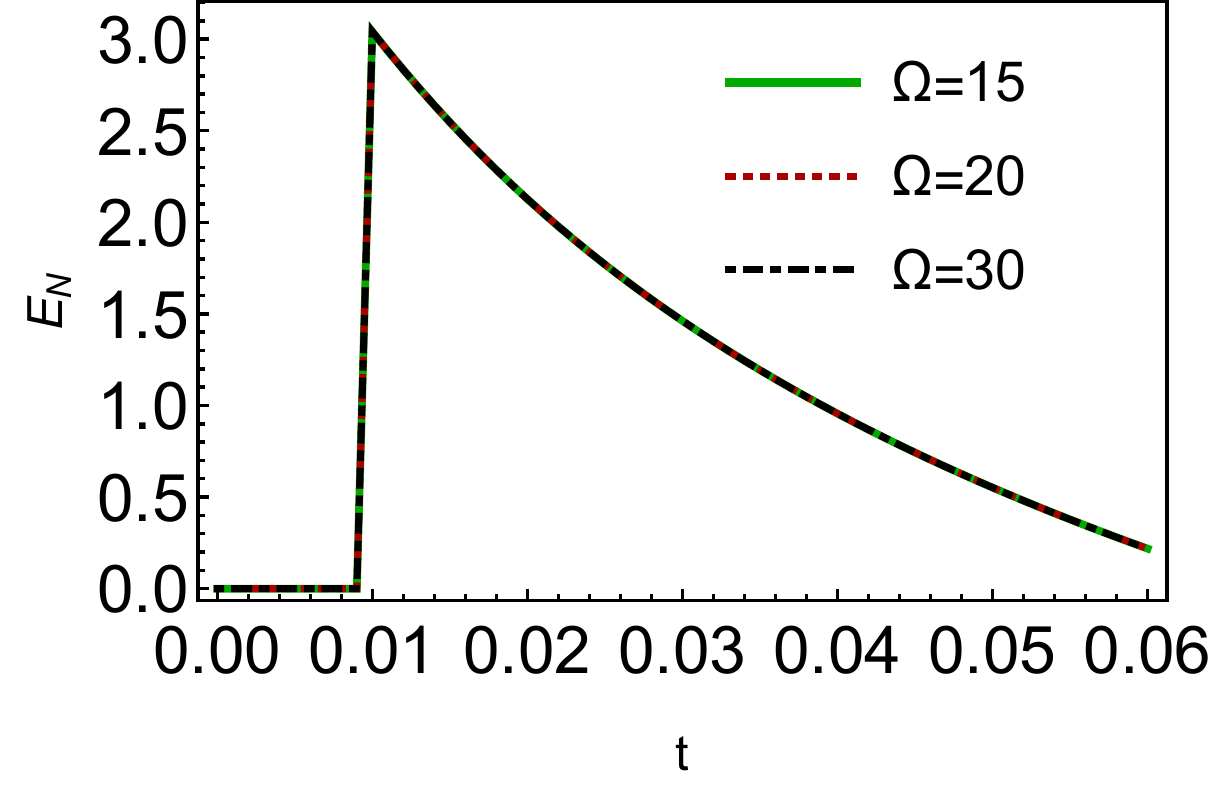}
\end{center}
\vspace{-0.5cm}
 \caption{\label{f:nega1}
Logarithmic negativity $E_{\mathcal{N}}$ of the micromechanical mirrors for the cavity drive lasers' for thermal phonon numbers, $n_1=n_2=50$ and thermal photon numbers $N_1=N_2=1$  as a function of $t$ at constant ${\rm \Omega_{p}/\gamma}=0.018$ with $\Omega=15$ (solid green line), $\Omega=20$ (dotted red line), and $\Omega=30$ (dotdashed black line). Here ${\rm g_{1}}={\rm g_{2}}=2\pi\times4{\rm MHz}$, ${\rm \kappa_{1}}={\rm \kappa_{2}}=2\pi\times215{\rm kHz}$, and assuming that all atoms are initially in their excited state $|\psi_{A}(0)\rangle=\ket{c}$, that is, for the value of the parameter $\eta=1$.} 
\end{figure}

\begin{figure}[tb]
\vspace{0.5cm}
\begin{center}
\includegraphics[width=0.995\columnwidth]{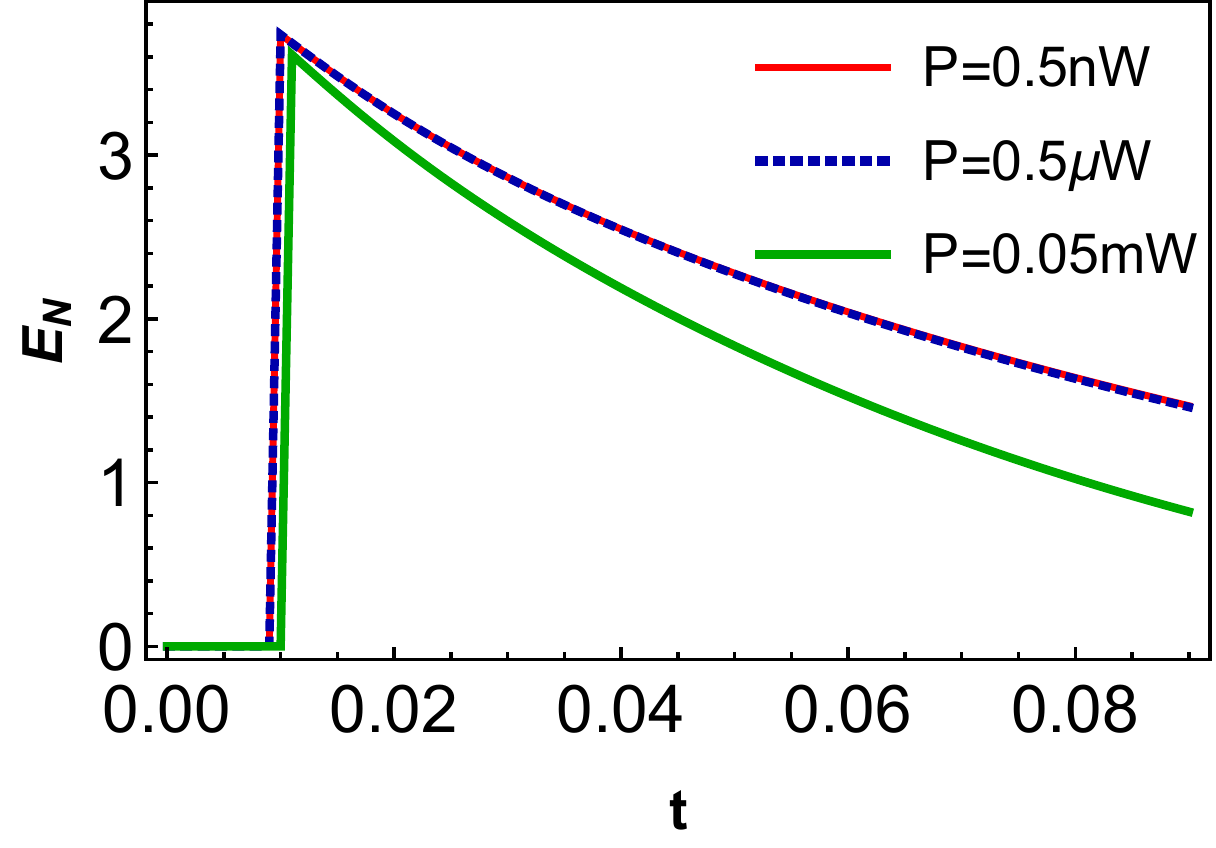}
\end{center}
\vspace{-0.5cm}
 \caption{\label{f:nega2}
Logarithmic negativity $E_{\mathcal{N}}$ of the micromechanical mirrors for thermal phonon numbers $n_1=n_2=5$ and thermal photon numbers $N_1=N_2=1$  as a function of $t$ at constant ${\rm \Omega_{p}/\gamma}=0.018$ and ${\rm \Omega/\gamma}=5$ with $\text{P}_1=\text{P}_2=\text{P}=0.5\text{nW}$ (solid red line), 
$\text{P}_1=\text{P}_2=\text{P}=0.05\mu\text{W}$ (dotted blue line), and $\text{P}_1=\text{P}_2=\text{P}=0.5\text{mW}$ (solid green line). Here ${\rm g_{1}}={\rm g_{2}}=2\pi\times4{\rm MHz}$, ${\rm \kappa_{1}}={\rm \kappa_{2}}=2\pi\times215{\rm kHz}$, and assuming that all atoms are initially in their excited state $|\psi_{A}(0)\rangle=\ket{c}$, i.e., for the value of the parameter $\eta=1$.} 
\end{figure}

\begin{figure}[tb]
\vspace{0.5cm}
\begin{center}
\includegraphics[width=0.995\columnwidth]{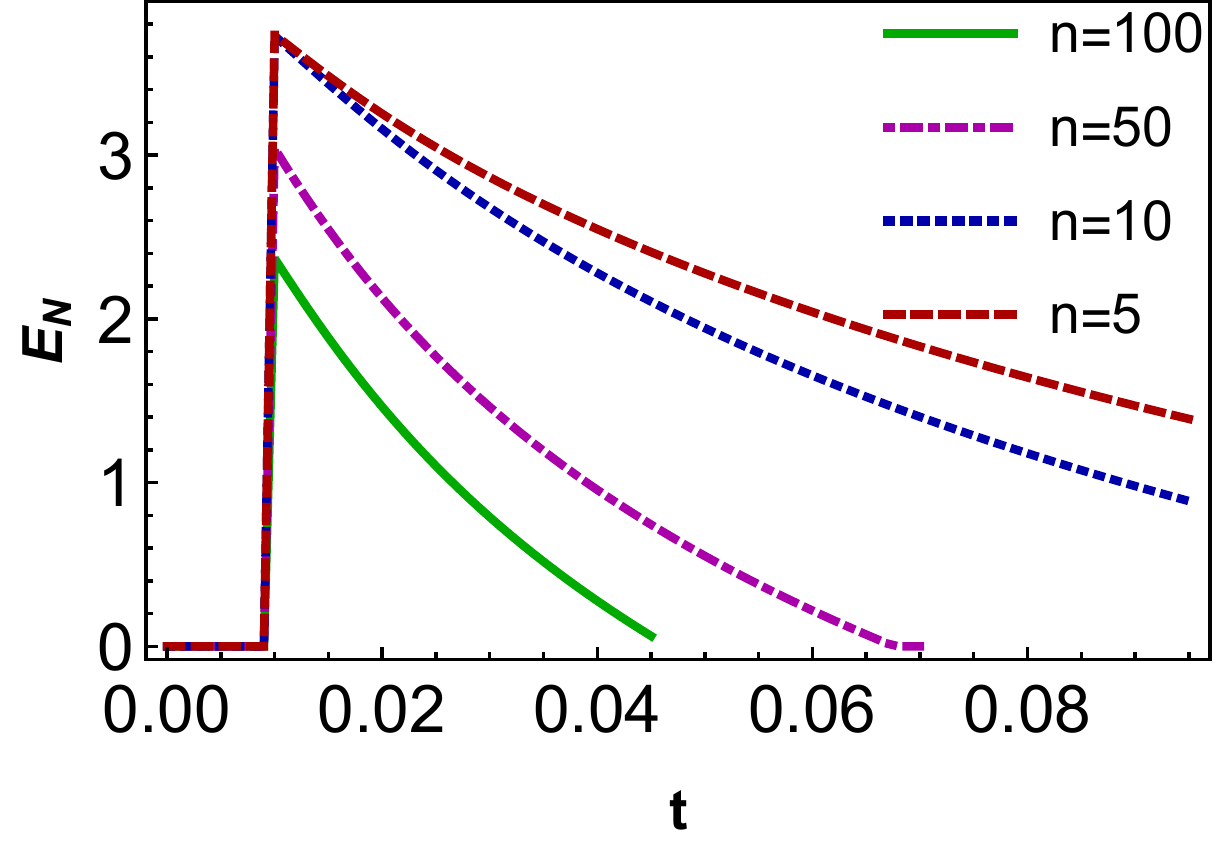}
\end{center}
\vspace{-0.5cm}
 \caption{\label{f:nega3}
Logarithmic negativity $E_{\mathcal{N}}$ of the micromechanical mirrors for thermal photon numbers $N_1=N_2=1$  as a function of $t$ at constant ${\rm \Omega_{p}/\gamma}=0.018$ and ${\rm \Omega/\gamma}=5$ for fixed cavity drive laser at $\text{P}_1=\text{P}_2=\text{P}=0.02\text{nW}$ with $n_1=n_2=n=100$ (solid green curve), $n_1=n_2=n=50$ (dotdashed magenta line), $n_1=n_2=n=10 $ (blue dotted line) and $n_1=n_2=n=5$ (red dashed line). See text and the above figures for other parameters.} 
\end{figure}

In Fig. \ref{f:nega1}, we plot the degree of entanglement of the two movable mirrors tunable as a function of time $t$ for different $\Omega$ at fixed input laser powers $P$, thermal noises $n$ and thermal photon numbers $N$. We consider the standard case where the case of symmetric mechanical damping ($\gamma_1=\gamma_2=\gamma$), symmetric thermal occupation of the mechanical baths ($n_1=n_2=n$) and symmetric thermal photon numbers ($N_1=N_2=N$).  This allows fast numerical results for the time dependent second moments to be evaluated. The assumption of equal thermal occupations is reasonable for most experimental situations, while it turns out that our results are not sensitive to unequal mechanical damping rates provided that they are both small. We observe that the amount of entanglement decreases with time and it is clear that the amount of entanglement are the same provided that with increasing the external driving field, $\Omega$ and saturated for all values of the external driving field. In turn this analysis shows that the generated entanglement can be controlled by adjusting
experimental conditions, particularly the external driving field $\Omega$.
To see the effect of the cavity-driving laser powers $P$ on the output entanglement, we plot the time dependence of the entanglement for various cavity-driving laser powers when all atoms are initially in their excited state $|\psi_{A}(0)\rangle=\ket{c}$, that is, for the value of the parameter $\eta=1$ in Fig. \ref{f:nega2} for thermal phonon numbers $n_1=n_2=n=5$ and thermal photon numbers $N_1=N_2=N=1$ . We observe that the degree of entanglement $E_{\mathcal{N}}$ increases and persists for longer time when the cavity-driving power $P$ decreases and the two movable mirrors are entangled for a wide range of the drive laser powers and saturated (for $\text{P}<0.05\mu\text{W}$). This is due to the coupling of the cavity-field mode to a mirror.    
We next examine the effect of the thermal noise on the degree of entanglement. The degree of entanglement of the two movable mirrors, as a function of time with the external driving field held constant, are shown in Fig. \ref{f:nega3}. We observe in the figures that the degree of entanglement has a similar curve to the effects of the cavity drive lasers for a small input power $P$ and a small thermal noise $n$. We also see that the degree of entanglement for the mirrors is reduced with increasing temperature. We see that the critical time above which the logarithmic entanglement $E_{\mathcal{N}}$ disappears increases with decreasing phonon thermal numbers. This is reminiscent of entanglement sudden-death where it does not exponentially decay but goes to zero at a critical time \cite{yu,lin}.   

\begin{figure}[tb]
\vspace{0.5cm}
\begin{center}
\includegraphics[width=0.995\columnwidth]{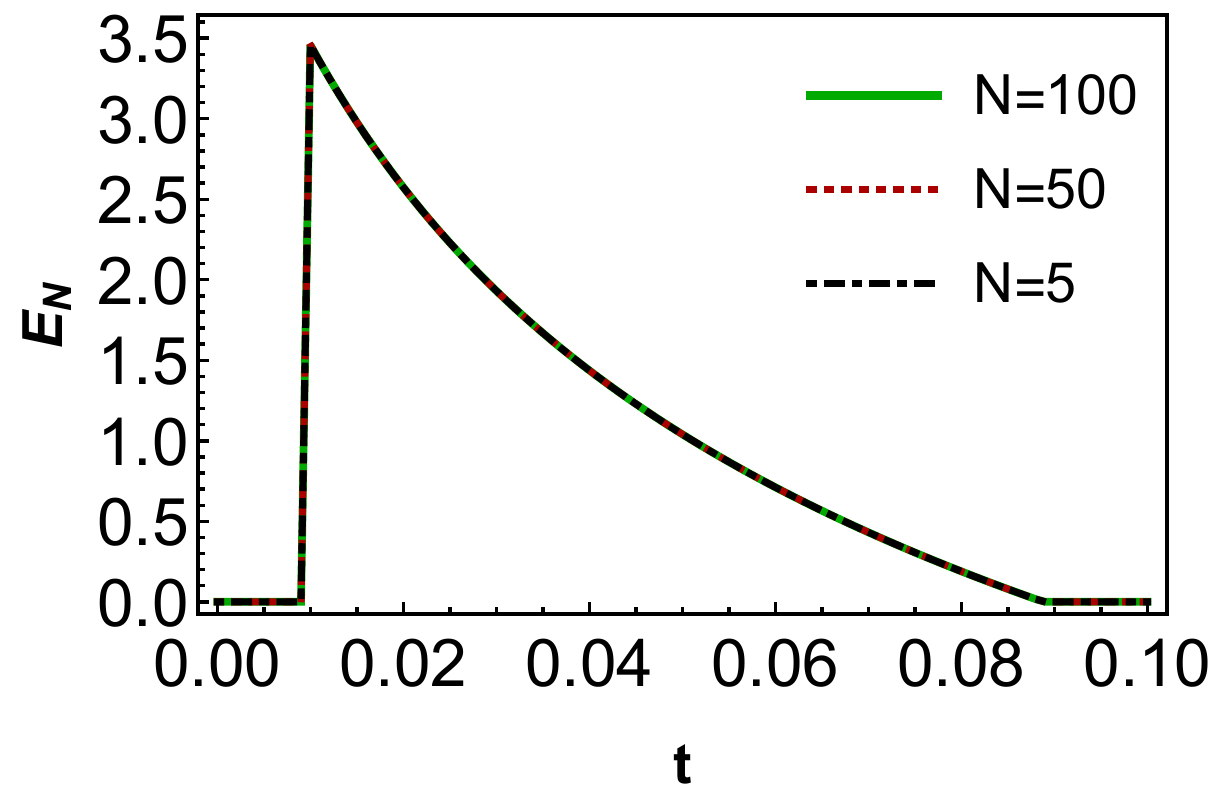}
\end{center}
\vspace{-0.5cm}
 \caption{\label{f:nega4}
Logarithmic negativity $E_{\mathcal{N}}$ of the micromechanical mirrors when the temperature of the thermal phonon bath is zero, $T=0K$ ($n_1=n_2=n=0$) as a function of $t$ at constant ${\rm \Omega_{p}/\gamma}=0.018$ and ${\rm \Omega/\gamma}=5$ for fixed cavity drive laser at $\text{P}_1=\text{P}_2=\text{P}=0.02\text{nW}$ with $N_1=N_2=N=100$ (solid green line), $N_1=N_2=N=50$ (dotted red line), and $N_1=N_2=N=5$ (dotdashed black line) for the value of the parameter $\eta=-1$. See text and the above figures for other parameters.} 
\end{figure}

Finally, we address the environmental temperature dependence of the two movable mirrors, as shown in Fig. \ref{f:nega4}. We see that at zero thermal phonon temperature and fixed cavity drive power $P_1=P_2=P=0.02\text{nW}$, the entanglement decreases irrespective of the number of thermal photons and persists for longer time. Moreover, we see that the critical time above which the entanglement disappears remains the same with varying thermal photons.

 \section{Conclusion}\label{sec:7}
 
We have presented a study of the optical bistability and entanglement between two mechanical oscillators coupled to the cavity modes of a two-mode laser via optical radiation pressure with realistic parameters. In stark contrast to the usual S-shaped bistability observed in single-mode dispersive optomechanical coupling, we have found that the optical intensities of the two cavity modes exhibit bistabilities for all large values of the detuning, due to the parametric amplification-type coupling induced by the two-photon coherence.   We have also investigated the entanglement of the movable mirror by exploiting the intermode correlation induced by the two-photon coherence. We have here focused on the dynamics of the quantum fluctuations of the mirror. We have shown that strong mirror-mirror entanglement can be created in the adiabatic regime. The degree of entanglement $E_{\mathcal{N}}$ is significant for a low thermal noise $n$ and a low cavity-driving laser powers $P$. The entanglement is supported by direct numerical calculations for realistic parameters. Our results suggest that for experimentally accessible parameters \cite{Gr,Ar}, macroscopic entanglement for two movable mirrors can be achieved with current technology and have important implications for quantum logic gates based on EIT schemes \cite{Feizpour}.

\section*{Acknowledgments}
B. T.  gratefully acknowledges numerous discussions with Eyob A. Sete.   B. T. is supported by the Shanghai Research Challenge Fund; New York University Global Seed Grants for Collaborative Research; National Natural Science Foundation of China (61571301); the Thousand Talents Program for Distinguished Young Scholars (D1210036A); and the NSFC Research Fund for International Young Scientists (11650110425); NYU-ECNU Institute of Physics at NYU Shanghai; the Science and Technology Commission of Shanghai Municipality (17ZR1443600); and the China Science and Technology Exchange Center (NGA-16-001) and by Khalifa University Internal Research Fund (8431000004).

\appendix

\section* {APPENDIX: COEFFICIENTS IN THE MASTER EQUATION (\ref{eq:master_final})}
\label{app:master}

In this section we derive the coefficients that appear in the master equation (\ref{eq:master_final}) relevant for our system. We follow an identical procedure to 
Ref. \cite{eyob15} to obtain the master equation (\ref{eq:master_eq}) 
\cite{Scu-book97,louisell,kassahun}.  The next step is to derive the conditioned density operators $\rho_{ac}={\bra a}\rho_{AR}\ket{c}$ and $\rho_{db}=\bra d\rho_{AR}{\ket b}$ and their complex conjugate as appeared in Eq. (\ref{eq:master_eq}). In the same way as Ref. \cite{eyob15}, we obtain for the matrix elements
\begin{equation}
\label{eq:reducedf} 
 \frac{d}{dt}\rho_{lk}(t)=r_a\rho_{lk}^{(0)}\rho-i{\bra l}\left[\mathcal {V}_1,\rho_{AR}(t)\right]\ket k-\gamma_{lk}\rho_{lk}.
 \end{equation}
Including the spontaneous emission and dephasing process, we can thus determine the equations for $\rho_{ac}$ and $\rho_{db}$ using (\ref{eq:reducedf}) gives (\ref{eq:rhoac&bd} and  (\ref{eq:rhoac&bd1}).    
  
To study the dynamics of our system, we make use of the linear approximation by keeping terms only up to the second order in the coupling constants, $\g (j=1,2)$ and consider all orders in the Rabi frequencies in the master equation. The nature of the linear approximation means that does not have saturation effects in the linear amplification regime. This is justified because the coupling constant of the two quantum fields are small as compared to other system parameters occurs on which dominates in the time evolution \cite{kiffner}. The zeroth-order equations of motion for $\rho_{aa}$, $\rho_{bb}$, $\rho_{cc}$, $\rho_{dd}$, $\rho_{ad}$, $\rho_{bc}$, $\rho_{ab}$, and $\rho_{cd}$ in the coupling constant are
\begin{align}
\label{eq:zeroth} 
\dot \rho_{aa}(t)&=r_{a}\rho_{aa}^{(0)}\rho-i\Omega(\rho_{ba}-\rho_{ab})-\gamma_{a}\rho_{aa}\\
\dot \rho_{bb}(t)&=r_{a}\rho_{bb}^{(0)}\rho-i\Omega(\rho_{ab}-\rho_{ba})-\gamma_{b}\rho_{bb}\\
\dot \rho_{cc}(t)&=r_{a}\rho_{cc}^{(0)}\rho-i\Omega_{p}(\rho_{dc}-\rho_{cd})-\gamma_{c}\rho_{cc}\\
\dot \rho_{dd}(t)&=r_{a}\rho_{dd}^{(0)}\rho-i\Omega_{p}(\rho_{cd}-\rho_{dc})-\gamma_{d}\rho_{dd}\\
\dot \rho_{ad}(t)&=r_{a}\rho_{ad}^{(0)}\rho-(\gamma_{ad}+i(\Delta_c-\Delta_1))\rho_{ad}\notag\\
&-i(\Omega\rho_{bd}-\Omega_{p}\rho_{ac})\\
\dot \rho_{bc}(t)&=r_{a}\rho_{bc}^{(0)}\rho-(\gamma_{bc}-i(\Delta_c-\Delta_2))\rho_{bc}\notag\\
&-i(\Omega\rho_{ac}-\Omega_{p}\rho_{bd})\\
\dot \rho_{ab}(t)&=r_{a}\rho_{ab}^{(0)}\rho-(\gamma_{ab}+i\Delta_c)\rho_{ab}
-i\Omega(\rho_{bb}-\rho_{aa})\\
\dot \rho_{cd}(t)&=r_{a}\rho_{cd}^{(0)}\rho-(\gamma_{cd}+i(\Delta_c-\Delta_1-\Delta_2))\rho_{cd}\notag\\
&-i\Omega_p(\rho_{dd}-\rho_{cc})\label{eq:zeroth1} 
 \end{align} 
in which $\gamma_{j}(j=a,b,c,d)$ are the $j$th atomic-level spontaneous emission rates and $\gamma_{ij}$ are the dephasing rates. We next need to apply the good-cavity limit where the cavity damping rate is much smaller than the dephasing and spontaneous emission rates. In this limit, the cavity mode variables slowly varying than the atomic variables, and thus the atomic variables converge to a steady state quickly. The steady state is found by setting the time derivatives in (\ref{eq:zeroth})-(\ref{eq:zeroth1}) to zero and the resulting algebraic equations can be solved exactly
\begin{equation}
\begin{aligned}
\label{eq:steady} 
\rho_{aa}&=\frac{r_a\rho}{d}  {Z_{aa}},\quad \rho_{bb}=\frac{r_a\rho}{d}  Z_{bb}\\
\rho_{ab}&=\frac{r_a\rho}{d} Z_{ab}\quad \rho_{cd}=\frac{r_a\rho_{cd}(0)}{d^\prime}  {Z_{cd}}\\
\rho_{cc}&=\frac{r_a\rho}{d^\prime} Z_{cc}\quad \rho_{dd}=\frac{r_a\rho}{d^\prime}Z_{dd}\\
Z_{aa}&=\left(2\Omega^{2}\gamma_{ab}+\gamma_{b}\left(\gamma_{ab}^{2}+\Delta_{c}^{2}\right)\right)\frac{\left(1-\eta\right)}{2}\\
Z_{ab}&=\frac{i\Omega\gamma_{b}\chi}{\gamma_{ab}+i\Delta_{c}}\frac{\left(1-\eta\right)}{2}\\
Z_{bb}&=\Omega^2\gamma_{ab}(1-\eta)\quad Z_{cc}=(2\Omega_{p}^2\gamma_{cd}+\gamma_{d}\chi^{\prime})\frac{(1+\eta)}{2}\\
Z_{dd}&=\Omega_{p}^{2}\gamma_{cd}\left(1+\eta\right)\\
Z_{cd}&=\frac{i\Omega_{p}\gamma_{d}\chi^\prime}{\gamma_{cd}+i\left(\Delta_{c}-\Delta_{2}-\Delta_{1}\right)}\frac{\left(1+\eta\right)}{2}
\end {aligned}
\end{equation}
with $d=2\Omega^2\gamma_{ab}(\gamma_a+\gamma_b)+\chi\gamma_b\gamma_a$, $\chi=\gamma^{2}_{ab}+\Delta_{c}^2$, $\chi^{\prime}=\gamma_{cd}^2+(\Delta_c-\Delta_2-\Delta_1)^2$, $d^{\prime}=2\Omega_{p}^2\gamma_{cd}(\gamma_c+\gamma_d)+\gamma_c\gamma_d\chi^{\prime}$. It proves to be more convenient to introduce a new parameter $\eta\in\left[-1,1\right]$ defined by $\rho_{aa}^{(0)}=\frac{1-\eta}{2}$, so that in view of the fact that $\rho_{aa}^{(0)}+\rho_{cc}^{(0)}=1$ and the initial coherence takes the form $\rho_{ac}{(0)}=\frac{1}{2}(1-\eta)^{1/2}$. The equations of motion, (\ref{eq:rhoac&bd}) and (\ref{eq:rhoac&bd1}), can be solved using the adiabatic approximation and the expressions for $ \rho_{ac}$, $\rho_{cc}$, 
$\rho_{bc}$, $\rho_{ad}$, $\rho_{bd}$ $\rho_{bb}$, $\rho_{dd}$, and $\rho_{ad}$, so that (\ref{eq:coupled1}) are obtained.  Here we define:
 \begin{align}
\xi_1=\frac{g^2_{1} A}{B^2-AD}\frac{r_a}{d^{\prime}}Z_{dd}\quad\xi_2=\frac{g^2_1 A}{B^2-AD}\frac{r_a}{d}Z_{bb}\\
\xi_3=\frac{g^2_{2} D}{B^2-AD}\frac{r_a}{d^{\prime}}Z_{cc}\quad \xi_4=\frac{g^2_2 D}{B^2-AD}\frac{r_a}{d}Z_{aa}\\
\xi_5=\frac{g_1g_2 B}{B^2-AD}\frac{r_a}{d}Z_{aa}\quad\xi_6=\frac{g_1g_2 B}{B^2-AD}\frac{r_a}{d^{\prime}}Z_{cc}\\
\xi_7=\frac{g_1g_2 B}{B^2-AD}\frac{r_a}{d}Z_{bb}\quad\xi_8=\frac{g_1g_2 B}{B^2-AD}\frac{r_a}{d^{\prime}}Z_{dd}
 \end{align}
 where 
 \begin{align}
 A&=-(\gamma_{ac}+i\Delta_2)-\frac{\Omega^2}{\gamma_{bc}-i(\Delta_c-\Delta_2)}-\frac{\Omega_{p}^2}{\gamma_{ad}+i(\Delta_c-\Delta_1)}\\
 B&=\frac{\Omega\Omega_{p}}{\gamma_{bc}-i(\Delta_c-\Delta_2)}+\frac{\Omega\Omega_{p}}{\gamma_{ad}+i(\Delta_c-\Delta_1)}\\
 D&=-(\gamma_{bd}-i\Delta_1)-\frac{\Omega^2}{\gamma_{ad}+i(\Delta_c-\Delta_1)}-\frac{\Omega_{p}^2}{\gamma_{bc}-i(\Delta_c-\Delta_2)}.
  \end{align}
Substituting (\ref{eq:coupled1}) into (\ref{eq:master_eq}) gives the master equation  (\ref{eq:master_final}) for the cavity modes. 


\begin{thebibliography}{99}
\bibitem{Bose} S. Bose, K. Jacobs, and P. L. Knight, \pra {\bf 56}, 4175 (1997).

\bibitem{Mancini} S. Mancini, V. I. Man'ko, and P. Tombesi, \pra {\bf55}, 3042 (1997). 

\bibitem{Genes} C. Genes, A. Mari, P. Tombesi, and D. Vitali, \pra {\bf78}, 032316 (2008).
\bibitem{Vitaliprl}D. Vitali, S. Gigan, A. Ferreira, H. R. B{\"o}hm, P. Tombesi, A. Guerreiro, V. Vedral, A. Zeilinger, and M. Aspelmeyer, \prl {\bf98}, 030405 (2007).

\bibitem{Mari09} A.  Mari  and  J.  Eisert,  \prl {\bf103}, 213603 (2009).
\bibitem{Jie} Jie-Qiao Liao and C. K. Law, \pra {\bf83}, 033820 (2011).
\bibitem{Genes08} C. Genes, D. Vitali, and P. Tombesi,\pra {\bf77}, 050307 (R) (2008). 
\bibitem{Ian} H. Ian, Z. R. Gong, Yu-xi Liu, C. P. Sun, and F. Nori, \pra {\bf78}, 013824 (2008); K. Hammerer, M. Aspelmeyer, E. S. Polzik, and P. Zoller, \prl {\bf102}, 020501 (2009).
\bibitem{Paz} Juan Pablo Paz and Augusto J. Roncaglia, \prl {\bf100}, 220401 (2008).
\bibitem{Liao} Jie-Qiao Liao, Qin-Qin Wu, and Franco Nori, \pra {\bf 89}, 014302 (2014).
\bibitem{Michael} Michael J. Hartmann and Martin B. Plenio, \prl {\bf101}, 200503 (2008).
\bibitem{Ling}  L. Zhou, Y. Han, J. Jing, and W. Zhang, \pra {\bf83}, 052117 (2011).
\bibitem{Ge13a} W. Ge, M. Al-Amri, H. Nha, and M. S. Zubairy,
\pra{\bf88}, 052301 (2013).
\bibitem{Ge13} W. Ge, M. Al-Amri, H. Nha, and M. S. Zubairy, \pra {\bf88}, 022338 (2013).

\bibitem{Mancini02} Stefano Mancini, Vittorio Giovannetti, David Vitali, and Paolo Tombesi, \prl {\bf88}, 120401 (2002).
\bibitem{Auer} Adrian Auer and Guido Burkard, \prb {\bf85}, 235140 (2012).
\bibitem{Mauro14} M. F. Pereira and I. A. Faragai, Optics express 22 (3), 3439-3446 (2014).
\bibitem{Mauro15} M. F. Pereira, Opt Quant Electron 47, 815-820 (2015).

\bibitem{Mauro16} M. F. Pereira, Appl. Phys. Lett. 109, 222102 (2016).
\bibitem{Mauro17}  M. F. Pereira, J. P. Zubelli, D. Winge, A. Wacker  A. S. Rodrigues, V. Anfertev and V. Vaks, \prb {\bf96}, 045306 (2017).



\bibitem{Xiong} H. Xiong, M. O. Scully, and M. S. Zubairy, \prl {\bf94}, 023601 (2005).
\bibitem{Kiffner} M. Kiffner, M. S. Zubairy, J. Evers, and C. H. Keitel, \pra {\bf75}, 033816 (2007).

\bibitem{Qamar} S. Qamar, M. Al-Amri, S. Qamar, and M. S. Zubairy, \pra {\bf80}, 033818 (2009).
\bibitem{Qamar1} S. Qamar, M. Al-Amri, and M. S. Zubairy, \pra {\bf79}, 013831 (2009).

\bibitem{eyob15} Eyob A. Sete and H. Eleuch, J. Opt. Soc. Am. B {\bf32}, 971-982 (2015). 
\bibitem{Tredicucci} A. Tredicucci, Y. Chen, V. Pellegrini, M. Borger, and F. Bassani, \pra {\bf54}, 3493-3498 (1996).
\bibitem{Dorsel} A. Dorsel, J. D. McCullen, P. Meystre, E. Vignes, and H. Walther, \prl {\bf51}, 1550 (1983).
\bibitem{Aspelmeyer} M. Aspelmeyer, T. J. Kippenberg, and F. Marquardt, Rev. Mod. Phys. {\bf86}, 1391-1452 (2014).

\bibitem{Scu-book97} M. O. Scully and M. S. Zubairy, \textit{Quantum Optics} (Cambridge University Press, 1997).

\bibitem{eyob11} Eyob A. Sete, \pra {\bf 84}, 063808 (2011). 
\bibitem{walls} D. F. Walls and G. J. Milburn, \textit{Quantum Optics} (Springer, 2008). 
\bibitem{eyob07} E. Alebachew, Opt. Commun. {\bf280}, 133-141 (2007).
\bibitem{eyob07a} E. Alebachew,  \pra {\bf76}, 023808 (2007).
\bibitem{eyob08} E. A. Sete, Opt. Commun. {\bf281}, 6124-6129 (2008). 
 \bibitem{Ge} Wenchao Ge and M Suhail Zubairy, Phys. Scr. {\bf90}, 074015 (2015). 
   \bibitem{Cohen}   Claude Cohen-Tannoudji, Jacques Dupont-Roc, Gilbert Grynberg, \textit{Atom-Photon Interactions: Basic Processes and Applications} (Wiley, New York, 2004).
 \bibitem{Hald} J  Hald and E S Polzik, J. Opt. B: Quantum Semiclass. Opt. {\bf3} S83 (2001). 
  
  
  
\bibitem{Jiang} C. Jiang, H. X. Liu, Y. S. Cui, X. W. Li, G. B. Chen, and X. M. Shuai, \pra {\bf88}, 055801 (2013).
\bibitem{Gr} S. Simon Gr\"{o}blacher, Klemens Hammerer, Michael R. Vanner, and Markus Aspelmeyer, Nature {\bf 460}, 724-727 (2009).
\bibitem{Ar} O. Arcizet, P.-F. Cohadon, T. Briant, M. Pinard, and A. Heidmann, Nature {\bf444}, 71-74 (2006).

\bibitem{Clerk}  A. A. Clerk, M. H. Devoret, S. M. Girvin, F. Marquardt, and R. J. Schoelkopf, Rev. Mod. Phys. {\bf82}, 1155 (2010).


 
 
 
 


\bibitem{Rist} D. Rist\'e, M. Dukalski, C. A. Watson, G. de Lange, M. J. Tiggelman, Ya. M. Blanter, K. W. Lehnert, R. N. Schouten,
and L. DiCarlo, Nature (London) {\bf502}, 350 (2013).
\bibitem{Palo} T. A. Palomaki, J. D. Teufel, R. W. Simmonds, and K. W. Lehnert, Science {\bf342}, 710 (2013).
\bibitem{Wang} Y.-D. Wang and A. A. Clerk, \prl {\bf110}, 253601 (2013); L. Tian, \textit{ibid}. {\bf110}, 233602
(2013); Z.-Q. Yin and Y.-J. Han, \pra {\bf79}, 024301(2009); M. C. Kuzyk, S. J. van Enk, and H. Wang, \textit{ibid}. {\bf88}, 062341 (2013); C. Joshi, J. Larson, M. Jonson, E. Andersson, and P. \"{O}hberg, \textit{ibid}. {\bf85}, 033805 (2012).

\bibitem{Doherty} A. C. Doherty and K. Jacobs, \pra {\bf60} (4), 2700 (1999). 

\bibitem{Warwick} W. P. Bowen and G. J. Milburn, \textit{Quantum Optomechanics} (CRC Press, Taylor \& Francis Group, LLC, 2016).

\bibitem{Pinard} M. Pinard, A. Dantan, D. Vitali, O. Arcizet, T. Briant, and A. Heidmann, Europhys. Lett. {\bf72}, 747-753 (2005).

\bibitem{Mari} A.  Mari,  A.  Farace,  N.  Didier,  V.  Giovannetti,  and  R. Fazio, \prl {\bf111}, 103605 (2013).

\bibitem{Vidal} G. Vidal and R. F. Werner, \pra {\bf 65} (3), 032314 (2002).
\bibitem{Adesso} G. Adesso, A. Serafini, and F. Illuminati, \pra {\bf70}, 022318 (2004).

\bibitem{Ferraro} A.  Ferraro,  S.  Olivares,  and  M.~G.~A.~Paris, \textit{Gaussian states in continuous variable quantum information} (Bibliopolis, Napoli, 2005).
\bibitem{yu} Ting Yu,, J. H. Eberly, Science {\bf323}, 598?601 (2009).
\bibitem{lin} Qing Lin, Bing He, R. Ghobadi, and Christoph Simon,  \pra {\bf 90}, 022309 (2014).

\bibitem{Feizpour}  Amir Feizpour, Greg Dmochowski, and Aephraim M. Steinberg, \pra {\bf93}, 013834 (2016).
\bibitem{louisell} W. H. Louisell, \textit{Quantum Statistical Properties of Radiation} (Wiley, New York, 1973).
\bibitem{kassahun} F. Kassahun, \textit{Fundamentals of Quantum Optics} (Lulu, Raleigh, NC, 2008).


\end{thebibliography}
\end{document}